\documentclass[a4paper,11pt]{article}

\usepackage{jheppub} 

\usepackage[T1]{fontenc} 
\usepackage{cancel}
\usepackage{hyperref}

\usepackage{amsmath}
\usepackage{relsize}

\usepackage{tikz,environ}
\usetikzlibrary{snakes}
\usetikzlibrary{decorations}
\usetikzlibrary{trees}
\usetikzlibrary{decorations.pathmorphing}
\usetikzlibrary{decorations.markings}
\usetikzlibrary{external}
\usetikzlibrary{intersections}
\usetikzlibrary{shapes,arrows}
\usetikzlibrary{arrows.meta}
\usetikzlibrary{calc}
\usetikzlibrary{shapes.misc}
\usetikzlibrary{decorations.text}
\usetikzlibrary{backgrounds}
\usetikzlibrary{fadings}
\usetikzlibrary{tikzmark,calc,arrows,shapes,decorations.pathreplacing}

\usetikzlibrary{shapes.geometric,positioning}

\tikzset{
	graviton/.style={decorate,line width=0.1mm, decoration={snake,amplitude=.3mm, segment length=0.8mm}},
	photon/.style={decorate, decoration={snake,amplitude=.4mm, segment length=2mm}, draw=red},
	scalar/.style={postaction={decorate},
	},
	massive/.style={postaction={decorate},
		line width=0.75mm,
	},
	massless/.style={postaction={decorate},
	},
	masslessWithDot/.style={postaction={decorate},
		decoration={
			markings,
			mark=at position 0.5 with {\fill circle (2pt);}}
	},
	massiveWithDot/.style={postaction={decorate},
		line width=0.5mm,
		decoration={
			markings,
			mark=at position 0.5 with {\fill circle (2pt);}}
	},
	massiveWithArrow/.style={postaction={decorate},
		line width=0.75mm,
		decoration={
			markings,
			mark=at position 0.5 with {\arrow{latex}}}
	},
	massiveLin/.style={postaction={decorate},
		double,
		thick,
		fill=white
	},
	massivePhi/.style={postaction={decorate},
		line width=0.75mm,
		dashed
	},
	masslessPhi/.style={postaction={decorate},
		dashed
	},
	unitaryCut/.style={postaction={draw,densely dashed,blue,thin},
		line width = 0.2cm,white
	},
	gluon/.style={decorate, draw=magenta,
		decoration={coil,amplitude=4pt, segment length=5pt}},
	partial ellipse/.style args={#1:#2:#3}{
		insert path={+ (#1:#3) arc (#1:#2:#3)}
	},
	cross/.style={cross out, draw=black, minimum size=2*(#1-\pgflinewidth), inner sep=0pt, outer sep=0pt},
	branchCut/.style={postaction={decorate},
		snake=zigzag,
		decoration = {snake=zigzag,segment length = 2mm, amplitude = 2mm}	
	}
	cross/.default={1pt}
}

\colorlet{mred}{black!30!red}
\colorlet{mgreen}{black!30!green}
\colorlet{mblue}{black!30!blue}
\colorlet{morange}{blue!70!red}

\newcommand{\realpartNLONoInsertion}{\begin{tikzpicture}[scale=1]
		\coordinate (e1) at (-1,1);
		\coordinate (e4) at (-1,-1);
		\coordinate (v1) at (-0.5,1);
            \coordinate (v3) at (0.5, 1);
            \coordinate (v2) at (-0.5,-1);
            \coordinate (v4) at (0.5,-1);
		\coordinate (e2) at (1,-1);	
		\coordinate (e3) at (1,1);
     
            \draw[massive] (e1) -- (v1)-- (v3) -- (e3);
            \draw[massive] (e4) -- (v2) -- (v4) -- (e2);
            \draw[photon] (v1) -- (v2);
            \draw[photon] (v3) -- (v4);

            \node[left] at (-0.5, 0){$ l $};
            \node[right] at (0.5, 0){$ q-l $};
            
            \draw[line width=0.3mm, dashed, name path=rung4, blue] (0, 1.25) to (0, -1.25);
\end{tikzpicture}}

\graphicspath{{figures/}}
\usepackage[export]{adjustbox}

\usepackage{color}

\begin{document} 

\title{\boldmath One-Loop Observables to Higher Order in Spin}



\author{Juan Pablo Gatica}


\affiliation{Mani L Bhaumik Institute for Theoretical Physics,\\UCLA Department of Physics and Astronomy, Los Angeles, CA 90095 USA}

\emailAdd{jpgatica3541@g.ucla.edu}

\abstract{We study observables in the scattering of classical, spinning objects using the KMOC formalism. In particular, we derive formulas to higher order in spin and one loop $\mathcal{O}(G^2)$ for the spin kick and momentum impulse. Our derivation method is agnostic to the choice of theory or special conditions, such as the spin supplementary condition (SSC); we only rely on the generic structure of long-range scattering amplitudes of non-transverse, massive spinning fields in the classical limit. We check these formulas for the case of gravity and agree with previous results from the eikonal formalism after imposing a SSC.

}

\maketitle
\flushbottom

\newpage

\section{Introduction}

\hspace{0.6cm}The discovery of gravitational waves~\cite{LIGOScientific:2016aoc, lIGOScientific:2017vwq} has ushered in a new way of observing our universe. Nearly a decade into the age of gravitational wave astronomy, we have found a plethora of merger events whose contents include black holes, neutron stars, and combinations of the two~\cite{O3BCATALOG}.  We also have a number of future facilities to look forward to, such as LISA and the Einstein Telescope, which will be much more precise and wider in scope in their observations~\cite{Punturo:2010zz, LISA:2017pwj, Reitze:2019iox}. Clearly based on the current experimental landscape (and its foreseeable future), we need to accurately understand the underlying system contributing to these merger events, namely the gravitational two-body problem.

Since its inception, we have been calculating general-relativistic corrections to Newtonian gravity~\cite{Einstein:1916vd}. With the direct observation of gravitational waves, a renewed focus has been put into obtaining ever more precise corrections to the dynamics of the two-body system. The Post-Newtonian (PN) expansion, which describes the non-relativistic, weak-field limit of the two-body system, has had a rich and successful history in calculating these corrections~\cite{Barker:1970zr, Barker:1975ae, Kidder:1992fr, Kidder:1995zr, Blanchet:1998vx, Tagoshi:2000zg,
Porto:2005ac, Faye:2006gx, Blanchet:2006gy, Damour:2007nc, Steinhoff:2007mb, Levi:2008nh,
Steinhoff:2008zr, Steinhoff:2008ji, Marsat:2012fn, Hergt:2010pa,
 Porto:2010tr,  Levi:2010zu,  Porto:2010zg, Levi:2011eq,
 Porto:2012as,  Hergt:2012zx,  Bohe:2012mr,  Hartung:2013dza,  Marsat:2013wwa,  Levi:2014gsa,
Vaidya:2014kza, Bohe:2015ana,  Bini:2017pee,  Siemonsen:2017yux,
 Porto:2006bt,  Porto:2007tt,  Porto:2008tb,  Porto:2008jj,  Levi:2014sba,
Levi:2015msa, 
Levi:2015uxa,  Levi:2015ixa,  Levi:2016ofk,   Levi:2019kgk,  Levi:2020lfn,
 Levi:2020kvb,  Levi:2020uwu,  Kim:2021rfj,  Maia:2017gxn,  Maia:2017yok,  Cho:2021mqw,  Cho:2022syn,
 Kim:2022pou,  Mandal:2022nty, Kim:2022bwv, Mandal:2022ufb, Levi:2022dqm,  Levi:2022rrq}, with the state of the art reaching 6PN~\cite{Bini:2020hmy}. Simultaneously, the Post-Minkowskian (PM) expansion, which is the relativistic, weak-field limit of the two-body system, has also had a long history~\cite{Bini:2017xzy, Bini:2018ywr, 
Maybee:2019jus,   Guevara:2019fsj,
  Chung:2020rrz,   Guevara:2017csg,   Vines:2018gqi,  Damgaard:2019lfh,
Aoude:2020onz,
Vines:2017hyw,
Guevara:2018wpp,   Chung:2018kqs,
Chung:2019duq,
Bern:2020buy,
Kosmopoulos:2021zoq,
Liu:2021zxr,
Aoude:2021oqj,
Jakobsen:2021lvp,
Jakobsen:2021zvh,
Chen:2021kxt,
Chen:2022clh,
Cristofoli:2021jas,
Chiodaroli:2021eug,
Cangemi:2022abk,
Cangemi:2022bew,
Haddad:2021znf,
Aoude:2022trd,
Menezes:2022tcs,
Bern:2022kto,
Alessio:2022kwv,
Alessio:2023kgf,
Bjerrum-Bohr:2023jau,
Damgaard:2022jem,
Haddad:2023ylx,
Aoude:2023vdk,
Jakobsen:2023ndj,
Jakobsen:2023hig,
Heissenberg:2023uvo,
Bianchi:2023lrg},
with the state of the art calculation being the 4PM correction~\cite{Bini:2024ijq, Bini:2024tft, Dlapa:2024cje, Jakobsen:2023pvx, Jakobsen:2023hig, Jing:2023okh, Damgaard:2023ttc, Jakobsen:2023ndj, Bern:2022jvn, Dlapa:2022lmu, Khalil:2022ylj, Bern:2021yeh, Dlapa:2021npj, Bern:2021dqo}. Understanding both of these expansions to high precision is paramount to our ability to effectively detect gravitational-wave events.

More recently, scattering amplitude methods have been applied to help push these expansion to higher orders~\cite{Bern:2024vqs, Parra-Martinez:2020dzs, Herrmann:2021tct, Smirnov:2023yhb, Laporta:2000dsw, Bern:2010ue, Bern:2019prr, Bern:2019crd, Monteiro:2014cda, Dlapa:2022wdu, Badger:2023eqz, Edison:2023qvg}.  In particular, these methods have been applied to higher-order corrections to numerous two-body observables for the PM expansion including the momentum impulse, spin kick, scattering waveform, mass absorption, 
etc.~\cite{Jones:2023ugm, Bern:2023ity, Alaverdian:2024spu, Kosmopoulos:2021zoq, Gatica:2023iws, Maybee:2019jus, Jakobsen:2023hig, Aoude:2021oqj, Bohnenblust:2024hkw, Bautista:2024emt, Aoude:2024jxd, Aoude:2023dui, Brandhuber:2024bnz, Brandhuber:2024qdn, Kosower:2018adc, Aoki:2024boe, Buonanno:2024byg, Dlapa:2024cje, Bohnenblust:2023qmy, Brandhuber:2023hhl, Luna:2023uwd, Bini:2024rsy, Herderschee:2023fxh, Kosmopoulos:2023bwc, Cristofoli:2021vyo, Georgoudis:2023eke, Bern:2023zkg}. In this paper, we will be focusing on the Kosower-Maybee-O'Connell (KMOC) formalism~\cite{Kosower:2018adc}, which has provided a very effective dictionary that directly links scattering amplitudes to physical observables.

In the context of the PN and PM expansions, there has been a significant effort to incorporate spin effects into the two-body problem in recent years. Many different methods have appeared in the literature to directly calculate two-body observables from amplitudes including the worldline formalism~\cite{Haddad:2024ebn, Jakobsen:2023hig}, the eikonal-phase formalism~\cite{Luna:2023uwd, Heissenberg:2023uvo}, EFT Hamiltonians~\cite{Alaverdian:2024spu, Bern:2023ity}, and KMOC~\cite{Gatica:2023iws, Maybee:2019jus, Aoude:2023dui, Aoude:2021oqj, FebresCordero:2022jts}, among others. The way in which one takes into account spin effects in their amplitude is crucial when calculating spin corrections to observables. Refs.~\cite{Chung:2018kqs, Guevara:2018wpp, Aoude:2020onz, Arkani-Hamed:2017jhn} have shown that, up to quartic order in spin, spin effects on the Compton amplitude for massive spinning fields minimally coupled to gravity exponentiate in the classical limit; however, at higher orders in spin, this exponentiation breaks down. Ref.~\cite{Bautista:2022wjf} showed, by matching higher-spin Compton amplitudes to solutions to the Teukolsky equation, that there are corrections to the amplitude that break the exponentiation, which have been related to subtleties in the Teukolsky solution~\cite{Bautista:2024agp, Bautista:2023sdf}. These subtleties will not affect our derivations in this work because we are checking our formulas up to quadratic order in spin. That being said, our derivation will not rely on this truncation and, in fact, we will be using properties of higher-spin fields for our calculation.

While a lot of work has been done using fixed-spin representations of massive fields to model scattering bodies~\cite{ Luna:2023uwd, Maybee:2019jus, Aoude:2023dui, Aoude:2021oqj, FebresCordero:2022jts}, the calculations in Refs.~\cite{Alaverdian:2024spu, Bern:2023ity, Gatica:2023iws} use non-transverse massive spinning fields, which allow for lower-spin states to propagate in their amplitudes, and with allowed transitions between them. This difference in models effectively implies a difference in choice of spin supplementary condition (SSC) and spin-vector-magnitude conservation: the former enforces it while the latter does not.  Ref.~\cite{Bern:2023ity} showed that one can go from SSC-violating to SSC-enforcing observables by properly tuning the Wilson coefficients in their QFT. This relationship implies that systems with the SSC imposed can be thought of as subsets of those that allow for SSC-violating degrees of freedom. In contrast to the non-transverse formalism, the fixed-spin formalism would have to introduce auxiliary fields at higher orders in spin that would result in complicated spin-$s$ propagators, such as those used in the massive higher-spin literature~\cite{Singh:1974qz, Lindwasser:2023dcv, Lindwasser:2023zwo, Bern:2023ity}. In this paper, we will use the simplifications allowed by SSC-violating degrees of freedom to calculate higher-spin corrections to the momentum impulse and spin kick.

The exponentiated S-matrix approach to calculate PM observables also has a rich history, with the use of the radial action~\cite{Akpinar:2024meg, Bern:2021yeh, Bern:2024adl, Gonzo:2024zxo, Dlapa:2024cje, Wilson-Gerow:2023syq, Bern:2021xze} and eikonal phase~\cite{Alaverdian:2024spu, Luna:2023uwd, Gatica:2023iws, Alessio:2024wmz, Bianchi:2023lrg, DiVecchia:2022nna, Bellazzini:2022wzv} providing very useful semi-classical approaches. While much of the work using these methods has been done for non-spinning matter, recent work has incorporated spin degrees of freedom to their formalisms. Based on their results using an EFT Hamiltonian, a formula was reverse engineered in Ref.~\cite{Bern:2020buy} that directly related the eikonal phase to the one-loop correction of the momentum impulse and spin kick. These eikonal formulas have the nice property of only having simple operations acting on the eikonal phase such as differentiation and commutation. In Refs.~\cite{Gatica:2023iws, Luna:2023uwd}, eikonal formulas were derived, where the former used non-transverse massive spinning fields to calculate up to linear order in spin, while the latter used fixed-spin fields and provided an expression the was verified up to quadratic order in spin. 

In this paper, we expand upon the work in Ref.~\cite{Gatica:2023iws} and generalize it to higher orders in spin. We use the KMOC formalism and take advantage of the properties of non-transverse massive spinning fields to simplify the derivation. In Section 2, we review the key elements for performing the derivation. In Section 3, we discuss properties of the amplitudes we will be working with. In Section 4 and 5, we carry out the derivation for the momentum impulse and spin kick eikonal formulas, respectively. In Section 6, we point out an underlying pattern in our eikonal formulas and discuss this in the context of the existing literature. In Section 7, we summarize our results and consider possible new directions to elaborate on this work. We also provide appendices that clarify the details of our calculation and comparison.

\section{Review}
\hspace{0.6cm}We review the necessary building blocks for deriving the eikonal formulas for the momentum impulse and the spin kick. First, we will review the scaling arguments that will allow us to sensibly take a classical limit. Next, we briefly touch on the basic assumption of our massive higher-spin fields. Then, we briefly cover the KMOC formalism with spin included. Later, we present the special kinematics used to simplify the organization of the classical limit. Finally, we review the eikonal-phase formalism and recent efforts in deriving eikonal formulas. 

\subsection{The Classical Scaling with Spin}
\hspace{0.6cm}Taking the classical limit, in the context of two-body scattering amplitudes, amounts to determining the relevant length scales of the problem: there are three in the massive two-body context~\cite{Correia:2024jgr}. We must consider the de Broglie wavelength, $\lambda_{dB} = \hbar/|\mathbf{p}|$, where $\mathbf{p}$ is the three-momentum of either body, which is associated with the scale where the wave nature of the bodies becomes important; the impact parameter $b$, which is the macroscopic length of separation between the two bodies; and the Compton wavelength $\lambda_C = \hbar/m$ where $m$ is the mass of either body, which is associated with the scale at which QFT effects, such as particle production, become relevant. For our purposes, we take the long-range, relativistic, classical limit, 
\begin{equation}
    b \gg \lambda_{dB} \sim \lambda_{C},
\end{equation}
otherwise known as the eikonal limit~\cite{Correia:2024jgr}. 

For scattering amplitudes, it is more convenient to talk about this scaling in momentum space. To do this, we need the conjugate variable to the impact parameter, which is the momentum transfer $|q| \sim \hbar/|b|$. This implies the following scaling relation of our momenta,
\begin{equation}
    |q|\sim \hbar \ll m \sim |\mathbf{p}|.
\end{equation}

We can also relate the total angular momentum to the impact parameter by $|J| \sim |b| |\mathbf{p}|$, which implies $J \gg \hbar$ in the eikonal limit. Of course, we are interested in the spin angular momentum $S$ of the scattering; therefore, we will assume that when we distinguish $S$ from the orbital angular momentum $L$, their magnitudes scale similarly, $|S|\sim |L| \gg \hbar$~\cite{Bern:2020buy}.

From now on, we will be working in natural units $\hbar=c=1$. To organize the scaling, we will use a small unitless parameter $\lambda \ll 1$ and have the following scaling in mind for our variables
\begin{equation}
    p \rightarrow p, \hspace{0.5cm} q \rightarrow \lambda q,\hspace{0.5cm} b \rightarrow \lambda^{-1} b,\hspace{0.5cm} S \rightarrow \lambda^{-1} S,
\end{equation}
when taking the eikonal limit. 

\subsection{Higher-Spin Fields}
\hspace{0.6cm}To model our scattering bodies, we will be following the work done in Refs.~\cite{Bern:2020buy, Bern:2023ity, Alaverdian:2024spu}, which use symmetric, non-transverse, massive higher-spin fields; we will refer to these as general-spin fields. By choosing general-spin fields, we allow for lower-spin states to propagate in our amplitudes, thus allowing for spin-transitions in our scattering events, which Ref.~\cite{Bern:2023ity} showed violates spin vector magnitude conservation while still conserving the spin tensor magnitude. Furthermore, our bodies no longer necessarily satisfy the covariant spin supplementary condition (SSC),
\begin{equation}
    p_{\mu}S^{\mu \nu}(p) = 0,
\end{equation}
therefore allowing for extra degrees of freedom in our QFT. When taking the classical limit of observables, Ref.~\cite{Bern:2023ity} showed that results that do not use a SSC and those that do are related by a choice in Wilson coefficients in the QFT. This implies that imposing the SSC on the general-spin fields eliminates these extra degrees of freedom in the classical limit, thus recovering the fixed-spin-$s$ result. This also implies that we can impose the SSC after we have calculated observables using general-spin fields and recover those that obey it; in other words, general-spin fields are a more general description of massive spinning bodies.  

A major advantage in using the general-spin fields is that we are able to construct a simple completeness relation for the spin-$s$ polarization tensors
\begin{equation}\label{completenessrelation}
    \sum_{a} \epsilon^{\mu(s)}_a (p) \epsilon^{*\,a}_{\nu(s)}(p) = \delta^{\mu(s)}_{\,\,\nu(s)},
\end{equation}
where we sum over the little group indices $a$ and $\delta^{\mu(s)}_{\,\,\nu(s)}$ is the appropriate Kronecker delta such that for a rank-$s$ tensor $\delta^{\mu(s)}_{\,\,\nu(s)} T^{\nu(s)} = T^{\mu(s)}$. 

We define the spin tensor using these polarization tensors
\begin{equation}\label{stensordef}
    [S^{\mu \nu}(p)]^{a'}_{\,\,a} \equiv \epsilon^{* \, a'}_{\alpha(s)}(p) \left(\mathit{M}^{\mu \nu}\right)^{\alpha(s)}_{\,\,\, \beta(s)} \epsilon^{\beta(s)}_{a}(p),
\end{equation}
where $\left(\mathit{M}^{\mu \nu}\right)^{\alpha(s)}_{\,\,\, \beta(s)}$ is the spin-$s$ representation of the Lorentz generator
\begin{equation}
    \left(\mathit{M}^{\mu \nu}\right)^{\alpha(s)}_{\,\,\, \beta(s)} = i s \, \delta^{[\mu}_{(\beta_1} \eta^{\nu] (\alpha_1} \delta^{\alpha_2}_{\beta_2} \cdots \delta^{\alpha_s)}_{\beta_s)},
\end{equation}
satisfying the Lorentz algebra
\begin{equation}
    [\mathit{M}^{\mu \nu}, \mathit{M}^{\rho \sigma}] = - i \left(\eta ^{\mu \rho} \mathit{M}^{\sigma \nu} + \eta ^{\nu \rho} \mathit{M}^{\mu \sigma} - \eta ^{\mu \sigma} \mathit{M}^{\rho \nu} - \eta ^{\nu \sigma} \mathit{M}^{\mu \rho}\right).
\end{equation}
Because the commutator in the Lorentz algebra reduces the number of Lorentz generators by one, for the purposes of classical scaling, we will take the commutator to scale as $\mathcal{O}(\lambda)$.

We define products of spin tensors in the same way as Refs.~\cite{Bern:2020buy, Bern:2023ity, Alaverdian:2024spu}, where for a product of $s$ spin tensors,
\begin{equation}
    \left[S^{\mu_1 \nu_1}(p) \cdots S^{\mu_s \nu_s}(p) \right]^{a'}_{a} = \frac{1}{s!}\epsilon^{*\,a'}(p)\left\{\mathit{M}^{\mu_1 \nu_1}, \left\{ \mathit{M}^{\mu_2 \nu_2}, \cdots ,\left\{\mathit{M}^{\mu_{s-1} \nu_{s-1}}, \mathit{M}^{\mu_s \nu_s}\right\} \cdots \right\}\right\} \epsilon_{a}(p). 
\end{equation}
For convenience we will use the shorthand,
\begin{equation}
    \left[S^{\mu(s) \nu(s)}(p)\right]^{a'}_{a} \equiv \left[S^{\mu_1 \nu_1}(p) \cdots S^{\mu_s \nu_s}(p) \right]^{a'}_{a}.
\end{equation}

\subsection{KMOC with Spin}
\hspace{0.6cm}The KMOC formalism calculates the change in some observable $\mathcal{O}$ by measuring the difference in the expectation value of some corresponding quantum operator $\mathbb{O}$ from the asymptotic past to the asymptotic future~\cite{Kosower:2018adc},
\begin{equation}\label{kmocbase}
    \Delta \mathcal{O} = \langle{\rm out}| \mathbb{O} |{\rm out} \rangle - \langle{\rm in}| \mathbb{O} |{\rm in} \rangle.
\end{equation}
We can relate the "out" states to the "in" state via the S-matrix, $S = 1 + i T$
\begin{equation}\label{vandr}
    \Delta \mathcal{O} = \langle{\Psi}|  i [\mathbb{O}, T] |{\Psi} \rangle + \langle{\Psi}|  T^{\dagger} [\mathbb{O}, T] |{\Psi} \rangle,
\end{equation}
where $|{\Psi} \rangle$ is the scattering wave packet. For a two-body wave packet we have 
\begin{equation}\label{wavepacket}
    | \Psi \rangle = \sum_{a_1, a_2} \int d \Phi(p_1)d \Phi(p_2) \, \phi(p_1)\phi(p_2) \, \xi^{a_1}\, \xi^{a_2} \, e^{i b_1 \cdot p_1}\, e^{i b_2 \cdot p_2} \, | p_1 p_2 ; a_1  a_2\rangle,
\end{equation}
where  $a_i$ are the little group indices of the scattering bodies, $\phi(p_i)$ are the wavefunctions associated with the bodies, $\xi^{a_i}$ are vectors in the little group, and we use momentum eigenstates~\cite{Kosower:2018adc, Maybee:2019jus, Luna:2023uwd, Cristofoli:2021jas}. We write the Lorentz invariant phase space as 
\begin{equation}
    \int d\Phi(p_i) \equiv \int \hat{d}^{D}p_i \, \hat{\delta}(p_i^2 - m_i^2),
\end{equation}
where $D = 4 - 2 \epsilon$, $\hat{d}^{D}p_i \equiv {d^{D}p_i}/{(2 \pi)^D}$, and $\hat{\delta}(p_i^2 - m_i^2) \equiv 2 \pi \theta(p_i^{0}) \delta(p_i^2 - m^2)$. Because we will be taking the classical limit, we will ignore the positive energy enforcing $\theta$-function from now on. We also assume a displacement $b_i$ for each particle in the wavepacket with respect to some arbitrary origin.

We can re-express the change in the observable for two-body scattering as 
\begin{equation}\label{kmocschem}
    \Delta \mathcal{O} = \prod_{i = 1}^{2} \, \sum_{a_i, a_i'}\int d \Phi(p_i) \, d \Phi(p_i') \,\phi^{*}(p_i')\phi(p_i) \, \xi^{*}_{a_i'} \, \xi^{a_i} \, e^{-i b_i \cdot( p_i'- p_i)} \, (\mathcal{I}^{\{a'_i\}}_{v \,\, \{a_i\}}  + \mathcal{I}^{\{a'_i\}}_{r \,\, \{a_i\}}),
\end{equation}
where
\begin{equation}
    \mathcal{I}^{\{a'_i\}}_{v \,\, \{a_i\}} = \langle{p_1' p_2';a_1' a_2'}|  i [\mathbb{O}, T] |{ p_1 p_2; a_1 a_2} \rangle, \hspace{0.5cm} \mathcal{I}^{\{a'_i\}}_{r \,\, \{a_i\}} = \langle{p_1' p_2';a_1' a_2'}|  T^{\dagger} [\mathbb{O}, T] |{p_1 p_2; a_1 a_2} \rangle,
\end{equation}
are known in the literature as the virtual kernel and real kernel, respectively. 

We define the amplitude as
\begin{align}\label{ampdef}
    \langle{p_1' p_2';a_1' a_2'}|&T|p_1 p_2; a_1 a_2\rangle = \hat{\delta}^{(4)}(p_1' + p_2' - p_1 - p_2) \left[\mathcal{A}\left(p_1, p_2, p_1', p_2', S(p_1), S(p_2)\right) \right]^{a_1' a_2'}_{\,\,a_1 a_2} \nonumber\\
    & \equiv \hat{\delta}^{(4)}(p_1' + p_2' - p_1 - p_2) \, \epsilon^{* \, a_1'}(p_1')\epsilon^{* \, a_2'}(p_2') \cdot \mathbb{A}(p_1, p_2, p_1', p_2') \cdot \epsilon_{a_1}(p_1)\epsilon_{a_2}(p_2),
\end{align}
where we leave products over representation indices implicit and $\mathbb{A}$ is the polarization-stripped amplitude, which helps clarify the little group dependence of the amplitude. Note that $\mathbb{A}$ only depends on the momenta and Lorentz generators. 

Recalling Eq. (\ref{stensordef}), we can see that all the little group dependence of the amplitude is in the spin tensor. Therefore, an important quantity we need to consider is the expectation value of the spin tensors 
\begin{equation}\label{coherentstatesum}
    \prod_{i = 1}^{2}\int d \Phi (p_i) |\phi(p_i)|^2 \sum_{a_i a_i'}  \xi^{*}_{a_i'} \,  \left[S^{\mu(s_i) \nu(s_i)}(p_i) \right]^{a_i'}_{\,\,a_i} \xi^{a_i} = \langle{S^{\mu(s_1) \nu(s_1)}_1 } \rangle\langle{S^{\mu(s_2) \nu(s_2)}_2 } \rangle.
\end{equation}
In the classical limit, following the arguments in Refs.~\cite{Kosower:2018adc, Maybee:2019jus, Luna:2023uwd, Cristofoli:2021jas}, the wave packets $\phi(p_i), \phi^{*}(p_i')$ sharply peak about their classical value. Similarly, when we consider the $\xi_{a_i}$ to be spin coherent states, the spin tensors will be sharply fixed to their classical value with minimal variance, which we have labeled as $S^{\mu \nu}_i$ for each spin tensor. Because all the little group information is contained in the spin tensor, we will ignore the little group indices for the amplitude under the understanding that the $S^{\mu(s_i)\nu(s_i)}_i$ are matrix valued in them; we will also ignore the angle brackets in the classical limit. 

With these simplifications in mind, when we take the classical limit we can express Eq. (\ref{kmocschem}) more simply
\begin{equation}\label{dO}
    \Delta \mathcal{O} = \int \hat{d}^{D}q \, \hat{\delta}(2 p_1 \cdot q + q^2) \, \hat{\delta}(2 p_2 \cdot q - q^2) \, e^{-i b \cdot q} \, (\mathcal{I}_v + \mathcal{I}_r) \equiv \int \cancel{D}q \, e^{-i b \cdot q} \, (\mathcal{I}_v + \mathcal{I}_r),
\end{equation}
where $q^{\mu} = p'^{\mu}_1 - p^{\mu}_1$ is the small momentum transfer conjugate to the impact parameter $b^{\mu} = b^{\mu}_2 - b^{\mu}_1$, and we have absorbed the $\delta$-functions into the the measure via the $\cancel{D}$ notation.

\subsection{Special Kinematics}
\hspace{0.6cm}When deriving the eikonal formulas, we want to explicitly show that our expressions uniformly scale classically; however the current kinematic set up makes this difficult. For example, in either of the momentum-conserving $\delta$-functions in Eq. (\ref{dO}), we can see an inhomogeneity in the classical scaling 
\begin{equation}
    \hat{\delta}(2 p_1 \cdot q + q^2) \rightarrow \hat{\delta}(2 \lambda p_1 \cdot q +  \lambda^2 q^2),
\end{equation}
which makes it difficult to keep track of the overall scaling of our integrands. A convenient special kinematic set up for the calculation of two-body scattering observables in the eikonal limit that simplifies this issue are the following~\cite{Parra-Martinez:2020dzs, Herrmann:2021tct},
\begin{equation}
    \bar{p}_1 = p_1 + q/2 ,\hspace{1cm} \bar{p}_2 = p_2 - q/2 ,\hspace{1cm} y = \frac{\bar{p}_1 \cdot \bar{p}_2}{\bar{m}_1 \bar{m}_2} ,\hspace{1cm} \bar{m}_i^2 = \bar{p}_i^2 = m_i^2 - q^2/4. 
\end{equation}
These kinematics simplify our $\delta$-functions to be $\hat{\delta}(2 \bar{p}_1 \cdot q )\hat{\delta}(2 \bar{p}_2 \cdot q )$, however the shift to special kinematics will have other effects on our observable integrands. 

\subsection{The Eikonal Phase}
\hspace{0.6cm}We will briefly review the eikonal-phase formalism and recent developments in this subject in the context of scattering observables. A detailed discussion on the eikonal phase can be found in~\cite{DiVecchia:2021bdo, DiVecchia:2022piu, DiVecchia:2023frv, Ajith:2024fna}. In impact parameter space, it has been observed that the two-body scattering amplitude exponentiates
\begin{equation}\label{eikexp}
    1 + i \, \mathcal{A}(b) = \left(1 +  i \Delta(b)\right)e^{i \delta(b)},
\end{equation}
where $\delta(b)$ is the eikonal phase, $\Delta(b)$ is a quantum remainder, and
\begin{equation}
    \mathcal{A}(b) = \int \cancel{D}q \, \mathcal{A}(q) \, e^{- i b \cdot q} \equiv \text{FT}[\mathcal{A}(q)],
\end{equation}
where we leave the dependence of the amplitude on other variables such as external momenta and spins implicit. In a recent work, the authors in Ref.~\cite{Du:2024rkf} provide a proof of Eq. (\ref{eikexp}) using the worldline formalism for non-spinning massive bodies. 

By expanding Eq.~(\ref{eikexp}) in powers of the coupling constant and in powers of classical scaling parameter $\lambda$, one can relate the amplitude order by order to the eikonal phase. Up to next to leading order in the coupling constant,
\begin{equation}\label{eiktoamps}
    \delta^{(1)}(b) = \text{FT}[\mathcal{A}^{(1)}(q)], \hspace{1cm} \delta^{(2)}(b) = \text{FT}[\text{Re}\, \mathcal{A}^{(2)} ], 
\end{equation}
where we assume that Eq.~(\ref{eiktoamps}) holds to all orders in the spin expansion and the superscript counts powers of the coupling constant. 

In Ref.~\cite{Bern:2020buy}, the authors calculated scattering observables for gravity by calculating amplitudes using general-spin QFT, relating this to an EFT potential and then used Hamilton's equations to calculate the momentum impulse and spin kick. This procedure has been used to calculate observables up to fifth power in spin at one loop~\cite{Bern:2022kto}. From their results, they were able to construct an ansatz that relates the eikonal phase to these observables via simple operations such as commutation or differentiation. In Refs.~\cite{Gatica:2023iws, Luna:2023uwd}, both works were able to derive an eikonal formula at one loop. In particular Ref.~\cite{Luna:2023uwd}, who used the eikonal formalism, was able to show that these one-loop eikonal formulas come from half variable shifts of their tree-level observables, suggesting an iterative pattern. 

While Ref.~\cite{Luna:2023uwd} was able to derive an one-loop eikonal formula that they validated to quadratic order in spin, they did so assuming a SSC and using fixed-spin states. In this work, we will refrain from imposing a SSC in our derivation in order to have as general a result as possible. As stated earlier, by allowing lower spin states to propagate in our fields we imply a violation of the SSC, which allows us to use simple relations such as Eq. (\ref{completenessrelation}) to derive the one-loop eikonal formulas. We will show that, after deriving the formulas, we can properly impose the SSC to recover the result found in Ref.~\cite{Luna:2023uwd}.

\section{General-Spin Amplitudes}
\hspace{0.6cm}When calculating the observables from general spin amplitudes we perform an expansion in the coupling constant in our QFT. We will organize this expansion in the following way
\begin{equation}
    \mathcal{A}(q, p_1, p_2, S_1, S_2) = g^2 \mathcal{A}^{(1)}(q, p_1, p_2, S_1, S_2) + g^4 \mathcal{A}^{(2)}(q, p_1, p_2, S_1, S_2) + \cdots
\end{equation}
where $g$ is some coupling constatnt and we will refer to $\mathcal{A}^{(1)}$ as the tree-level amplitude and $\mathcal{A}^{(2)}$ as the one-loop amplitude. There are basic structures of these amplitudes in the eikonal limit that are crucial to the calculation of the eikonal formulas.

\subsection{Tree-level Amplitudes}
\hspace{0.6cm}We know that the tree-level amplitude should have the following properties in the eikonal limit:
\begin{enumerate}
    \item Model elastic scattering;
    \item Lead to a Coulomb potential: $V(b) \propto 1/b$;
    \item Have the same classical scaling as the $s_1 = s_2 = 0$ case: $\mathcal{A}^{(1)} \rightarrow \mathcal{O}(\lambda^{-2}) \mathcal{A}^{(1)}$.
\end{enumerate}
With these properties we can construct an ansatz for the tree-level amplitude
\begin{equation}\label{treelvlansatz}
    \mathcal{A}^{(1)}(q, p_i, S_i) = e^{i q \cdot (\omega_1 - \omega_2)} i^{(s_1 + s_2)} \frac{q_{\mu(s_1)}q_{\rho(s_2)}}{q^2} S^{\mu(s_1) \nu(s_1)}_1 S^{\rho(s_2) \sigma(s_2)}_2 \Upsilon^{(1)}_{\nu(s_1) \sigma(s_2)}(p_1, p_2),
\end{equation}
where $q_{\mu(s_1)} = q_{\mu_1} q_{\mu_2} \cdots q_{\mu_{s_1}}$ and $\Upsilon^{(1)}_{\nu(s_1) \sigma(s_2)}(p_1, p_2)$ is some $s_1 +s_2$ ranked tensor at tree-level that is a function of the momenta. One can construct an ansatz for $\Upsilon$, however for the purposes of deriving the eikonal formulas this will not be necessary. We will include in the appendix the ansatz we used to perform our checks.

The exponential in Eq. (\ref{treelvlansatz}) comes from the exponentiation of the product of the polarization tensors~\cite{Gatica:2023iws, Bern:2020buy, Bern:2023ity, Chung:2018kqs}
\begin{equation}\label{polexp}
    \epsilon^{*}(p_1 + q) \cdot \epsilon(p_1) \, \epsilon^{*}(p_2 - q) \cdot \epsilon(p_2) = e^{i q \cdot (\omega_1 - \omega_2)},
\end{equation}
where $\omega^{\mu}_i(p_i) = S^{\mu \nu}_i (k_{i \, \nu} + p_{i \, \nu})/(p_i \cdot k_i + m_i^2)$. Here $k_i$ are the reference momenta of each scattering body, which is taken to be the rest frame. In the context of the KMOC formalism, we can absorb these exponentials into our impact parameter, such that we shift into a new position space coordinate  
\begin{equation}
    b_{cov} \equiv b - (\omega_1 - \omega_2), 
\end{equation}
which we will refer to as the covariant impact parameter. From now on we will be working with the covariant impact parameter and therefore we will ignore the exponential when considering Eq. (\ref{treelvlansatz}). For a discussion on changing between the canonical impact parameter, $b$, and $b_{\text{cov}}$, see Ref.~\cite{Bohnenblust:2024hkw}.

\subsection{One-Loop Amplitudes}
\hspace{0.6cm}The one-loop amplitude in the eikonal limit can be broken down into box, triangle, and bubble diagrams~\cite{Forde:2007mi, Parra-Martinez:2020dzs, Herrmann:2021tct}. Equivalently, we can decompose the one-loop amplitude into real and imaginary contributions. These two decompositions are related in the following way
\begin{align}
    &\mathcal{A}^{(2)} = \text{Re} \, \mathcal{A}^{(2)} + i \text{Im} \, \mathcal{A}^{(2)},\\
    &\text{Re} \, \mathcal{A}^{(2)} = \mathcal{A}^{(2)}_{\Delta} + \mathcal{A}^{(2)}_{\nabla} \rightarrow (\lambda^{-1}) \text{Re} \, \mathcal{A}^{(2)}, \\
    &\text{Im} \, \mathcal{A}^{(2)} = \mathcal{A}^{(2)}_{\text{Box}} + \mathcal{A}^{(2)}_{\text{xBox}} \rightarrow (\lambda^{-2}) \text{Im} \, \mathcal{A}^{(2)},
\end{align}
where we ignore the bubble contributions, since they do not contribute in the eikonal limit at this loop order. We have also highlighted the fact that the real and imaginary contributions have different classical scalings. The real contribution has the expected scaling to be a classical contribution. The imaginary contribution has a lower than expected scaling which would make it a super-classical or classically-singular contribution~\cite{Kosower:2018adc, Herrmann:2021tct, Bern:2021xze}. The classically-singular contribution is associated with the infrared divergence that arises from the sum of the box and cross-box diagrams; this divergence must vanish when calculating IR safe observables, such as the momentum impulse and spin kick, and we will show how this explicitly occurs when deriving the eikonal formulas. 

We can also construct an ansatz for $\text{Re} \, \mathcal{A}^{(2)}$ in a similar fashion to the tree-level ansatz,
\begin{align}\label{2pmansatz}
    \text{Re} \, \mathcal{A}^{(2)} =&  \frac{i^{(s_1 + s_2)} S^{\mu(s_1) \nu(s_1)}_1 S^{\rho(s_2) \sigma(s_2)}_2 }{\sqrt{-q^2}} \left\{q_{\mu(s_1)}q_{\rho(s_2)} \Upsilon^{(2)}_{\nu(s_1) \sigma(s_2)}\right.\nonumber\\
    & \left. + q^2 \left(q_{\mu(s_1-2)}q_{\rho(s_2)} \Upsilon^{(2)}_{\mu_{s_1} \mu_{s_1-1} \nu(s_1) \sigma(s_2)} + q_{\mu(s_1)}q_{\rho(s_2-2)} \Upsilon^{(2)}_{\nu(s_1) \rho_{s_2}\rho_{s_2-1} \sigma(s_2)}  \right. \right. \nonumber \\
    & \left. \left. \hspace{1cm}+ q_{\mu(s_1-1)}q_{\rho(s_2-1)} \Upsilon^{(2)}_{\mu_{s_1}\nu(s_1) \rho_{s_2}\sigma(s_2)} \right) + \cdots \right\},
\end{align}
where, in principle, we can continue including coefficients with higher powers of $q^2$ since these will not cancel the propagator. While Eq. (\ref{2pmansatz}) is not necessary to derive the eikonal formulas, it will be necessary when verifying them up to quadratic order in spin. We adopt the convention that $q_{\mu(-1)} = q_{\rho(-1)} = 0$ and $q_{\mu(0)} = q_{\rho(0)} = 1$. We also suppress the momentum dependence with the understanding the $\Upsilon^{(2)} \equiv \Upsilon^{(2)}(p_1, p_2)$.

\subsection{Unitarity with Higher Spin}
\hspace{0.6cm}As mentioned before, the imaginary part of the one-loop amplitude is the sum of the box and cross box diagrams. It is well known that this combination can be expressed as the on-shell product of tree-level amplitudes, which can be seen using unitarity and truncating to one-loop order
\begin{equation}\label{naiveIm}
    2 \, \text{Im} \, \mathcal{A}^{(2)} (q, p_1, p_2) = \int \cancel{D}l \, \mathcal{A}^{(1)}(q - l, p_1 +l, p_2 -l)  \, \mathcal{A}^{(1)}(l, p_1, p_2),
\end{equation}
where $l$ is the loop momentum and we have left the dependence on the spin tensors implicit. However, this product is misleading. Recall that when we take products of Lorentz generator we need to decompose terms into symmetric and anti-symmetric products. Stripping the amplitudes of their polarizations makes this clearer
\begin{align} \label{imnotsonaive}
    2 \, \text{Im} \, \mathbb{A}^{(2)} (q, p_1, p_2) &= \int \cancel{D}l \, \mathbb{A}^{(1)}(q - l, p_1 +l, p_2 - l)\,\mathbb{A}^{(1)}(l, p_1, p_2) \nonumber \\
    & = \frac{1}{2} \int \cancel{D}l \left(\left[\mathbb{A}^{(1)}(q - l, p_1 +l, p_2 - l), \mathbb{A}^{(1)}(l, p_1, p_2)\right] \right. \nonumber \\
    & \left. \hspace{2cm} + \left\{\mathbb{A}^{(1)}(q - l, p_1 +l, p_2 - l), \mathbb{A}^{(1)}(l, p_1, p_2)\right\}\right).
\end{align}
Now the anti-symmetric term has the same classical scaling as the real part of the one-loop amplitude i.e. it has been promoted from a classically-singular to a classical contribution. On the other hand, the symmetric product continues to be classically-singular. 

We still have yet to shift to special kinematics, which has the potential to promote the classical scaling of the amplitudes. Assuming that $q$ and $l$ have the same scaling, we can expand the amplitudes in small $\lambda$ (or equivalently soft internal momenta $q, l$) resulting in the following transformations to our tree-level amplitudes
\begin{align}
    &\mathbb{A}^{(1)}(l, p_1, p_2) \rightarrow \left(1 - \left(\frac{q - l}{2}\right) \cdot \left(\frac{\partial}{\partial \Bar{p}_1} - \frac{\partial}{\partial \Bar{p}_2}\right)\right)\mathbb{A}^{(1)}(l, \Bar{p}_1, \Bar{p}_2), \\
    &\mathbb{A}^{(1)}(q - l, p_1 + l, p_2 - l) \rightarrow \left(1 + \frac{l}{2} \cdot \left(\frac{\partial}{\partial \Bar{p}_1} - \frac{\partial}{\partial \Bar{p}_2}\right)\right)\mathbb{A}^{(1)}(q-l, \Bar{p}_1, \Bar{p}_2). 
\end{align}
This transformation does not affect the anti-symmetric term in Eq. (\ref{imnotsonaive}) because it already scales classically and any promotion would over-correct it to a quantum contribution. 

The shift to special kinematics also has another more subtle yet crucial effect. Recall that we are integrating over $\cancel{D}l$, which upon shifting to special kinematics becomes
\begin{equation}
    \cancel{D}l \rightarrow \hat{d}^{D}l \, \hat{\delta}(2 \bar{p}_1 \cdot l - q^2/2) \, \hat{\delta}(2 \bar{p}_2 \cdot l + q^2/2)
\end{equation}
where, while keeping in mind that we are only interested in long-range scattering, we make the replacement $q \cdot l \rightarrow q^2/2$ in order to avoid canceling any massless propagators. Once again we have a $\delta$-function that is inhomogenous in classical scaling, which allows us to expand it,
\begin{equation}
     \cancel{D}\,l \rightarrow \left(1 + \frac{q^2}{4} \left(\frac{\check{u}^{\alpha}_2}{\bar{m}_2} - \frac{\check{u}^{\alpha}_1}{\bar{m}_1}\right) \frac{\partial}{\partial l^{\alpha}}\right)\hat{\delta}\left(2 \bar{p}_1 \cdot l\right)\, \hat{\delta} \left(2 \bar{p}_2 \cdot l\right),
\end{equation}
where we truncate to the leading correction. Here $\check{u}_{1, 2}  \equiv (u_{1, 2} - y u_{2, 1})/(1-y^2)$, such that $\check{u}_i \cdot u_j = \delta_{i, j}$, where $u_i \equiv \bar{p}_i/\bar{m}_i$ is the classical velocity. We will eventually integrate by parts to remove the derivative from the $\delta$-functions, which will result in new classical contributions. Ref.~\cite{DeAngelis:2023lvf} used this procedure in a similar context. 

With the special kinematics applied, and restoring the polarization tensors, we have the following imaginary contribution to the one-loop amplitude
\begin{align}\label{ImSpecialKin}
    2 \,& \text{Im} \,  \mathcal{A}^{(2)}  (q, p_1, p_2)  \nonumber\\
    &= \int \cancel{D}l \left\{\mathcal{A}^{(1)}(l, \bar{p}_1, \bar{p}_2)\mathcal{A}^{(1)}(q-l, \bar{p}_1, \bar{p}_2) +\frac{1}{2}\left[\mathcal{A}^{(1)}(q -l, \bar{p}_1, \bar{p}_2), \mathcal{A}^{(1)}(l, \bar{p}_1, \bar{p}_2)\right] \right. \nonumber \\
    & \left. \hspace{1.75 cm} + \mathcal{A}^{(1)}(l, \bar{p}_1, \bar{p}_2)\, \frac{l}{2}\cdot \left(\frac{\partial}{\partial \bar{p}_1} - \frac{\partial}{\partial \bar{p}_2}\right)\mathcal{A}^{(1)}(q -l, \bar{p}_1, \bar{p}_2)  \nonumber \right. \\
    & \left. \hspace{2.0 cm} - \mathcal{A}^{(1)}(q -l, \bar{p}_1, \bar{p}_2) \frac{(q-l)}{2}\cdot \left(\frac{\partial}{\partial \bar{p}_1} - \frac{\partial}{\partial \bar{p}_2}\right) \mathcal{A}^{(1)}(l, \bar{p}_1, \bar{p}_2) \right\} \nonumber \\
    & \hspace{0.7cm}+ \left(\frac{\check{u}^{\alpha}_2}{\bar{m}_2} - \frac{\check{u}^{\alpha}_1}{\bar{m}_1}\right)\frac{q^2}{4} \int \hat{d}^{D}l \,  \mathcal{A}^{(1)}(l, \bar{p}_1, \bar{p}_2) \, \mathcal{A}^{(1)}(q-l, \bar{p}_1, \bar{p}_2)  \frac{\partial}{\partial l^{\alpha}} \left(\hat{\delta}\left(2 \bar{p}_1 \cdot l\right)\, \hat{\delta} \left(2 \bar{p}_2 \cdot l\right)\right),
\end{align}
where only the symmetric product in the first line, which is the naive unitarity relation, is still classically-singular.

In principle, $\text{Im} \, \mathcal{A}^{(2)}$ should obey what is known as the horizontal-flip symmetry~\cite{Jones:2023ugm, Herrmann:2021tct}, which is to say that the integrand in Eq. (\ref{ImSpecialKin}) should be invariant under the exchange of massless propagators $l \rightarrow q -l$. This is clear to see diagrammatically,
\begin{align}\label{Irdp}
   2 \, \text{Im} \, & \mathcal{A}^{(2)}  (q, p_1, p_2) = \int \hat{d}^{D} l \, \,  \vcenter{\hbox{\scalebox{1.10}{\realpartNLONoInsertion}}},
\end{align}
where the dashed line represents a unitarity cut. If we were to enforce this symmetry, only the symmetric product and last line of Eq. (\ref{ImSpecialKin}) would survive, since the other terms are parity-odd under this exchange. However, in the context of KMOC, we must first see how our operators affect the parity of the integrand before we start ruling out contributions by horizontal-flip symmetry.

\section{Momentum Impulse from KMOC with Spin}
\hspace{0.6cm}We will start by deriving the eikonal formula for the momentum impulse. Because our wavepackets are expressed in terms of momentum eigenstates, the momentum operator acts in the following way:
\begin{equation}
    \mathbb{P}^{\mu}_{1} | p_1, p_2 \rangle = p_1^{\mu} |p_1 p_2 \rangle.
\end{equation}
We will first calculate the virtual contribution followed by the real contribution. We will then reorganize the contributions to the momentum impulse into non-iteration and iteration pieces, focusing on the latter since it is more non-trivial.

\subsection{Virtual and Real Kernel Contribution}
\hspace{0.6cm}Applying the momentum operator, for the virtual contribution we get
\begin{equation}
    \langle{\Psi}|  i [\mathbb{P}^{\mu}, T] |{\Psi} \rangle = \int \cancel{D}q e^{- i b_{\text{cov}}\cdot q} \mathcal{I}^{\mu}_{v} = \int \cancel{D} q e^{- i b_{\text{cov}}\cdot q} i q^{\mu} \mathcal{A} (q, p_1, p_2).
\end{equation}
At tree level, this is the only contribution to the momentum impulse
\begin{align}
    \Delta^{(1)}p^{\mu}_1 &= \int \cancel{D} q e^{- i b_{\text{cov}}\cdot q} i q^{\mu} \mathcal{A}^{(1)} (q, p_1, p_2) \nonumber \\
    & = - \frac{\partial}{\partial (b_{\text{cov} \, \mu})_\perp} \int \cancel{D} q e^{- i b_{\text{cov}}\cdot q}  \mathcal{A}^{(1)} (q, p_1, p_2) = - \Pi^{\mu \nu} \frac{\partial \delta^{(1)}_{\text{cov}}}{\partial b^{\nu}_{\text{cov}}},
\end{align}
where we use the shorthand  $\delta_{\text{cov}} = \delta(b_{\text{cov}}, u_1, u_2, S_1, S_2)$ and
\begin{equation}
    \Pi^{\mu \nu} = \eta^{\mu \nu} - u^{\mu}_1 \check{u}^{\nu}_1 - u^{\mu}_2 \check{u}^{\nu}_2, \label{projector}
\end{equation}
is a projector that we are free to introduce due to the on-shell energy-conserving $\delta$-functions in $\cancel{D}q$. When exchanging $q^{\mu}$ for $i \partial/\partial b_{\text{cov} \, \mu}$, the projector preserves the on-shell condition upon taking derivatives. Anytime we are able to apply the on-shell conditions, we can introduce the projector in our integrand via the transformation $q^{\mu} \rightarrow \Pi^{\mu}_{\,\,\nu} q^{\nu}$.  Note that because the transfer momentum and impact parameter obey the same on-shell conditions ($b \cdot p_i = 0$), we can also freely make the exchange $b_{\text{cov}}^{\mu} \rightarrow \Pi^{\mu}_{\,\, \nu} b_{\text{cov}}^{\nu}$. This subtlety will become important once we move on to the one-loop case.

At tree level, we do not need to worry how switching to special kinematics will affect the scaling, since any promotions associated with the momentum shift would be an over-correction i.e., result in quantum contributions. Therefore, we can make the simple replacement $p_i \rightarrow \bar{p}_i$ and then take the classical value of the momenta $\bar{p}^{\mu}_i = \bar{m}_i u^{\mu}_i$.

At one loop, the virtual part of the momentum impulse can be broken down into contributions from the real and the imaginary part of the one-loop amplitude,
\begin{equation}
    \int \cancel{D} q e^{- i b_{\text{cov}}\cdot q} \left(i q^{\mu} \text{Re}\mathcal{A}^{(2)} (q, p_1, p_2) - q^{\mu} \text{Im}\, \mathcal{A}^{(2)}(q, p_1, p_2)\right) \equiv \Delta^{(2)}p^{\mu}_{1\,v, \text{Re}} + \Delta^{(2)}p^{\mu}_{1\,v, \text{Im}} .
\end{equation}
The contribution from $\Delta^{(2)}p^{\mu}_{1\,v, \text{Re}}$ will follow the same procedure in the tree-level case since it is already at the correct scaling to be considered classical,
\begin{align}
    \Delta^{(2)}p^{\mu}_{1\,v, \text{Re}} = \int \cancel{D} q e^{- i b_{\text{cov}}\cdot q} i q^{\mu} \text{Re}\mathcal{A}^{(2)} (q, \bar{p}_1, \bar{p}_2) = - \frac{\partial \delta^{(2)}_{\text{cov}}}{\partial (b_{\text{cov}\,\mu})_{\perp}}.
\end{align}

The real kernel only starts contributing at one loop. Plugging in for the momentum operator we get
\begin{align}
    \langle{\Psi}| T^{\dagger} [\mathbb{P}^{\mu}_{1}, T] |\Psi \rangle |_{\text{NLO}} &= \Delta^{(2)}p^{\mu}_{r}= \int \cancel{D}q e^{- i b_{\text{cov}} \cdot q} \mathcal{I}^{(2)\,\mu}_r \nonumber \\
    &= \int \cancel{D}q e^{- i b_{\text{cov}} \cdot q} \int \cancel{D}l \,l^{\mu} \mathcal{A}^{(1)}(q - l, p_1 + l, p_2 -l)\mathcal{A}^{(1)}(l, p_1, p_2).
\end{align}
As was the case for $\text{Im} \, \mathcal{A}^{(2)}$, we will need to strip the amplitudes of their polarizations, break down the product of amplitudes into symmetric and anti-symmetric parts, apply the momentum shift to special kinematics, and expand the shifted integrand. In the momentum impulse case, this procedure produces the same result as Eq. (\ref{ImSpecialKin}) except for the prefactor of $l^{\mu}$, which leads to a new organization of our momentum impulse contributions. 

\subsection{Iterative Contribution}
\hspace{0.6cm}We organize the remaining contributions to the momentum impulse, in the following way 
\begin{equation}\label{itorgdP}
    \Delta^{(2)}p^{\mu}_{it} \equiv  \Delta^{(2)}p^{\mu}_{r} + \Delta^{(2)}p^{\mu}_{1\,v, \text{Im}} = \Delta^{(2)}p^{\mu}_{it, *} + \Delta^{(2)}p^{\mu}_{it, \partial u} + \Delta^{(2)}p^{\mu}_{it, \partial l}.
\end{equation}
We will refer to Eq. (\ref{itorgdP}) as the iterative contribution since it is proportional to on-shell products of tree amplitudes.

The first term, $\Delta^{(2)}p^{\mu}_{it, *}$, is proportional to the symmetric and anti-symmetric product of the amplitudes 
\begin{align}
        \Delta^{(2)}p^{\mu}_{it, *} \equiv & \int \cancel{D}q e^{- i b_{\text{cov}} \cdot q} \int \cancel{D}l \left(\frac{l^{\mu} - (q^{\mu} - l^{\mu})}{2}\right) \nonumber  \\ 
    &\hspace{0.5cm}\left(\mathcal{A}^{(1)}(q-l, \bar{p}_1, \bar{p}_2) \mathcal{A}^{(1)}(l, \bar{p}_1, \bar{p}_2)  +\frac{1}{2}\left[\mathcal{A}^{(1)}(q -l, \bar{p}_1, \bar{p}_2), \mathcal{A}^{(1)}(l, \bar{p}_1, \bar{p}_2)\right] \right),  \label{dpitprod}
\end{align}
where the symmetric product still is classically-singular. To simplify Eq. (\ref{dpitprod}), we will enforce the horizontal-flip symmetry. Because the prefactor $(l^{\mu} - (q^{\mu} - l^{\mu}))/2$ is parity-odd under the exchange $l \rightarrow q -l$, the only term that will survive is the anti-symmetric product
\begin{equation}
     \Delta^{(2)}p^{\mu}_{it, *} = \int \cancel{D}q e^{- i b_{\text{cov}} \cdot q} \int \cancel{D}l  \left[\mathcal{A}^{(1)}(q -l, \bar{p}_1, \bar{p}_2), \frac{l^{\mu}}{2} \mathcal{A}^{(1)}(l, \bar{p}_1, \bar{p}_2)\right],
\end{equation}
where we exploited the overall horizontal-flip symmetry to write the above in a more compact form. By enforcing the horizontal-flip symmetry, we got rid of the only classically-singular term left in the calculation.

To express $\Delta^{(2)}p^{\mu}_{it, *}$ in terms of eikonal phases, we need to change the loop integral into a Fourier transform; to do this we perform the shift $q \rightarrow q + l$,
\begin{align}
    \Delta^{(2)}p^{\mu}_{it, *} & = \int \cancel{D}q e^{- i b_{\text{cov}} \cdot q} \int \cancel{D}l e^{- i b_{\text{cov}} \cdot l}  \left[\mathcal{A}^{(1)}(q, \bar{p}_1, \bar{p}_2), \frac{l^{\mu}}{2} \mathcal{A}^{(1)}(l, \bar{p}_1, \bar{p}_2)\right] \nonumber\\
    & = \frac{i}{2} \left[\delta^{(1)}_{\text{cov}}, \frac{\partial \delta^{(1)}_{\text{cov}}}{\partial (b_{\text{cov}\,\mu})_{\perp}}\right].
\end{align}

The second term in Eq. (\ref{itorgdP}), $\Delta^{(2)}p^{\mu}_{it, \partial u}$, is proportional to the external momentum derivatives 
\begin{align}
    \Delta^{(2)}p^{\mu}_{it, \partial u} \equiv &\int \cancel{D}q e^{- i b_{\text{cov}} \cdot q} \int \cancel{D}l \left(\frac{l^{\mu} - (q^{\mu} - l^{\mu})}{2}\right) \nonumber  \\ 
    &\hspace{0.5cm}\left( \mathcal{A}^{(1)}(l, \bar{p}_1, \bar{p}_2) \frac{l}{2}\cdot \left(\frac{\partial}{\partial \bar{p}_1}- \frac{\partial}{\partial \bar{p}_2}\right)\mathcal{A}^{(1)}(q -l, \bar{p}_1, \bar{p}_2) \nonumber \right.  \\
    & \left. \hspace{1.0cm} - \mathcal{A}^{(1)}(q -l, \bar{p}_1, \bar{p}_2) \frac{(q-l)}{2}\cdot \left(\frac{\partial}{\partial \bar{p}_1} - \frac{\partial}{\partial \bar{p}_2}\right) \mathcal{A}^{(1)}(l, \bar{p}_1, \bar{p}_2)  \right), \label{dpitdu}
\end{align}
For Eq. (\ref{dpitdu}), we will repeat the same procedure of enforcing the horizontal-flip symmetry, shifting the transfer momentum $q \rightarrow q + l$, and then re-expressing the amplitudes in terms of eikonal phases,
\begin{align}
    \Delta^{(2)}p^{\mu}_{it, \partial u} = -\frac{1}{2} \frac{\partial \delta^{(1)}_{\text{cov}}}{\partial (b_{\text{cov}\,\alpha})_{\perp}}& \frac{\overleftrightarrow{\partial}}{\partial (b_{\text{cov}\,\mu})_{\perp}}\left(\frac{1}{\bar{m}_1}\frac{\partial}{\partial u^{\alpha}_1}- \frac{1}{\bar{m}_2}\frac{\partial}{\partial u^{\alpha}_2}\right) \delta^{(1)}_{\text{cov}},
\end{align}
where 
\begin{equation}
    f(x) \frac{\overleftrightarrow{\partial}}{\partial x} g(x) \equiv \frac{\partial f(x)}{\partial x} g(x) - f(x)\frac{\partial g(x)}{\partial x}.
\end{equation}

The last term remaining, $\Delta^{(2)}p^{\mu}_{it, \partial l}$, comes from the expansion of the on-shell $\delta$-functions
\begin{align}
    \Delta^{(2)}p^{\mu}_{it, \partial l} \equiv &  \int \cancel{D}q e^{- i b_{\text{cov}} \cdot q} \int \hat{d}^{D}l \left(\frac{l^{\mu} - (q^{\mu} - l^{\mu})}{2}\right)\frac{q^2}{4} \left(\frac{\check{u}^{\alpha}_2}{\bar{m}_2} - \frac{\check{u}^{\alpha}_1}{\bar{m}_1}\right) \nonumber \\
    &\hspace{3.25cm} \mathcal{A}^{(1)}(l, \bar{p}_1, \bar{p}_2)\mathcal{A}^{(1)}(q-l, \bar{p}_1, \bar{p}_2)  \frac{\partial}{\partial l^{\alpha}}\left(\hat{\delta}\left(2 \bar{p}_1 \cdot l\right)\, \hat{\delta} \left(2 \bar{p}_2 \cdot l\right) \right) \label{dpitdl},
\end{align}
which will prove to be complicated to express in terms of eikonal phases. Say we integrate by parts to remove the derivative acting on the $\delta$-functions, carry out the derivatives, and repeat the procedure we used for the previous contributions
\begin{align}
    \Delta^{(2)}p^{\mu}_{it, \partial l} = & -\frac{1}{2} \left(\frac{\check{u}^{\mu}_2}{\bar{m}_2} - \frac{\check{u}^{\mu}_1}{\bar{m}_1}\right) \left(\frac{\partial \delta^{(1)}_{\text{cov}}}{\partial (b_{\text{cov}})_{\perp}}\right)^2 \nonumber \\
     & + \frac{1}{2} \left(\frac{\check{u}^{\alpha}_1}{\bar{m}_1} - \frac{\check{u}^{\alpha}_2}{\bar{m}_2}\right) \frac{\partial \delta^{(1)}_{\text{cov}}}{\partial (b_{\text{cov}\,\gamma})_{\perp}}  \frac{\overleftrightarrow{\partial}}{\partial (b_{\text{cov}\,\mu})_{\perp}} \int \cancel{D}l e^{- i b_{\text{cov}} \cdot l}  l_{\gamma} \frac{\partial}{\partial l^{\alpha}}\mathcal{A}^{(1)}(l, \bar{p}_1, \bar{p}_2). \label{troubletermdp}
\end{align}
We are still left with a term that does not neatly lend itself to being expressed as an eikonal phase. We must also be careful when applying the on-shell conditions in $\cancel{D}l$. Because the on-shell $\delta$-functions were being acted on by derivatives, we are not necessarily able to freely introduce the projector Eq. (\ref{projector}) into our amplitude prior to integrating by parts. As a result, when evaluating the integrand above, we must first apply the derivative on the amplitude, and only then we are free to introduce the on-shell projector, making this a correction to the cut condition of the internal massive lines in Eq. (\ref{Irdp}).  

To eventually express Eq. (\ref{troubletermdp}) fully in terms of eikonal phases, we plug in the ansatz for the tree-level general-spin amplitude Eq. ($\ref{treelvlansatz}$). After recognizing that the loop momentum derivative effectively replaces the projector $\Pi^{\mu \nu}$ with other variables, we prescribe the following replacement rule
\begin{equation}\label{replacementruledl}
    \int \cancel{D} l e^{- i b_{\text{cov}} \cdot l}  l_{\gamma}\frac{\partial}{\partial l^{\alpha}}\mathcal{A}^{(1)}(l, \bar{p}_1, \bar{p}_2) \rightarrow 2  \frac{\partial \delta^{(1)}_{\text{cov}}}{\partial \Pi^{\alpha \gamma}},
\end{equation}
where we define the derivative with respect to the projector as
\begin{equation}
    \frac{\partial \Pi_{\mu \nu}}{\partial \Pi_{\alpha \beta}} \equiv \frac{1}{2} \left(\delta^{\alpha}_{\mu} \delta^{\beta}_{\nu} + \delta^{\beta}_{\mu} \eta^{\alpha}_{\nu}\right),
\end{equation}
due to its symmetric nature. We must emphasize that taking the derivative with respect to the projector is more of a bookkeeping strategy that arrives at the desired expression; in the appendix, we provide an analysis of how we arrived at this resolution and an alternative way of expressing it. We also emphasize that if we had taken the derivative with the on-shell projector already applied Eq. (\ref{troubletermdp}) would have vanished. 

We now combine all of our contributions to the impulse to derive the higher-spin eikonal formula for the one-loop momentum impulse
\begin{align}\label{dPeikform}
    \Delta^{(2)}p^{\mu}_1 =& - \frac{\partial \delta^{(2)}_{\text{cov}}}{\partial (b_{\text{cov} \, \mu})_{\perp}} + \frac{i}{2} \left[ \delta^{(1)}_{\text{cov}}, \frac{\partial  \delta^{(1)}_{\text{cov}}}{\partial (b_{\text{cov} \, \mu})_{\perp}}\right]-\frac{1}{2}  \left(\frac{\check{u}_1^{\mu}}{\bar{m}_1} - \frac{\check{u}_2^{\mu}}{\bar{m}_2}\right) \left( \frac{\partial \delta^{(1)}_{\text{cov}}}{\partial (b_{\text{cov}})_{\perp}} \right)^2 \nonumber \\
    & + \frac{\partial \delta^{(1)}_{\text{cov}}}{\partial (b_{\text{cov} \, \alpha})_{\perp}}\frac{\overleftrightarrow{\partial}}{\partial (b_{\text{cov} \, \mu})_{\perp}} \left(\left(\frac{\check{u}_1^{\beta}}{\bar{m}_1} - \frac{\check{u}_2^{\beta}}{\bar{m}_2}\right) \frac{\partial}{\partial \Pi^{\alpha \beta}} - \frac{1}{2} \left(\frac{1}{\Bar{m}_1}\frac{\partial}{\partial u^{\alpha}_1} - \frac{1}{\Bar{m}_2}\frac{\partial}{\partial u^{\alpha}_2}\right) \right) \delta^{(1)}_{\text{cov}}.
\end{align}
We compared Eq. (\ref{dPeikform}) to the results in Ref.~\cite{Luna:2023uwd} up to quadratic order in spin and found full agreement. As another check we have verified that Eq. (\ref{dPeikform}) also satisfies momentum conservation, which can be easily seen if we make explicit the factors of $\Pi^{\mu \nu}$ in the derivative $\partial/\partial(b_{\text{cov}\,\mu})_{\perp}$ .

We can also rewrite Eq. (\ref{dPeikform}) in terms of the tree-level momentum impulse to simplify our expression,
\begin{align}\label{dPimpform}
    \Delta^{(2)}p^{\mu}_1=& - \frac{\partial \delta^{(2)}_{\text{cov}}}{\partial (b_{\text{cov} \, \mu})_{\perp}} - \frac{i}{2} \left[ \delta^{(1)}_{\text{cov}}, \Delta^{(1)}p^{\mu}_1 \right]-\frac{1}{2}  \left(\frac{\check{u}_1^{\mu}}{\bar{m}_1} - \frac{\check{u}_2^{\mu}}{\bar{m}_2}\right) \left( \Delta^{(1)} p_1 \right)^2 \nonumber \\
    & \hspace{0.25cm}+ \frac{1}{2}\Delta^{(1)}p_{1\,\alpha} \frac{\overleftrightarrow{\partial}}{\partial (b_{\text{cov} \, \mu})_{\perp}} \nabla^{\alpha}_{p_{\text{cm}}} \delta^{(1)}_{\text{cov}},
\end{align}
where we define
\begin{equation}
     \nabla^{\alpha}_{p_{\text{cm}}} \equiv  \left(\frac{1}{\Bar{m}_1}\frac{\partial}{\partial u_{1\,\alpha}} - \frac{1}{\Bar{m}_2}\frac{\partial}{\partial u_{2\,\alpha}}\right)- 2\left(\frac{\check{u}_{1\,\beta}}{\bar{m}_1} - \frac{\check{u}_{2\,\beta}}{\bar{m}_2}\right) \frac{\partial}{\partial \Pi_{\alpha \beta}} 
\end{equation}
as some derivative that respects the center of mass symmetry of the two-body system. We find that imposing a SSC removes the derivative part of the cut-correction contribution, Eq. (\ref{troubletermdp}), leading to the replacement $\nabla^{\alpha}_{p_{\text{cm}}} \rightarrow  \left(\frac{1}{\Bar{m}_1}\frac{\partial}{\partial u_{1\,\alpha}} - \frac{1}{\Bar{m}_2}\frac{\partial}{\partial u_{2\,\alpha}}\right)$. This is an interesting consequence of imposing the SSC and hints at some relation between these cut corrections and the choice of using non-transverse, massive spinning fields.

\section{Spin Kick from KMOC with Spin}
\hspace{0.6cm}In this section we will derive the higher-spin eikonal formula for the one-loop spin kick. Similarly to the momentum impulse case, we will begin by looking at the virtual and real kernels by applying the appropriate spin tensor operator 
\begin{equation}\label{exptS}
    \langle{p'_1 p_2'}| \mathbb{S}^{\mu \nu}_1 |p_1 p_2 \rangle  = \hat{\delta}_{\Phi}(p_1' - p_1) \hat{\delta}_{\Phi}(p_2' - p_2) \, \epsilon^{*}(p'_1)\cdot\mathit{M}^{\mu \nu} \cdot \epsilon(p_1),  
\end{equation}
where $\hat{\delta}_{\Phi}(p' - p)$ is the $\delta$-function for the Lorentz invariant phase space integral such that $\int d \Phi(p) \, f(p) \, \hat{\delta}_{\Phi}(p' - p) = f(p')$. We have rescaled our operator from the one in Ref.~\cite{Gatica:2023iws} in order to simplify our comparison to the literature. We will then reorganize our contributions to isolate the iteration terms and then simplify them into expressions proportional to products of eikonal phases. 

It will be more clear to work in polarization stripped amplitudes in this section, therefore we will introduce the shorthand $\epsilon \equiv \epsilon(p_1) \epsilon(p_2)$, $\epsilon^{*} \equiv \epsilon^{*}(p_1 + q)\epsilon^{*}(p_2- q)$. 

\subsection{Virtual and Real Kernel Contributions}
\hspace{0.6cm}Applying Eq. (\ref{exptS}), the virtual kernel reduces to   
\begin{align}
    \langle{\Psi}|  i [\mathbb{S}^{\mu \nu}_1, T] |{\Psi} \rangle = \int \cancel{D}q \, e^{- i b \cdot q} \, \mathcal{I}^{\mu \nu}_v &= -i\int \cancel{D}q \, e^{- i b \cdot q} \, \epsilon^{*} \cdot \left[\mathbb{A}(q, p_1, p_2), \mathit{M}^{\mu \nu} \right] \cdot \epsilon \nonumber\\
    & = -i\int \cancel{D}q \, e^{- i b_{\text{cov}} \cdot q} \, \left[\mathcal{A}(q, p_1, p_2), S^{\mu \nu}_1 \right], 
\end{align}
where it is understood that $\epsilon^{*}_{2'}, \epsilon_2$ do not contract with $\mathit{M}^{\mu \nu}$ and only interacts with the amplitude $\mathbb{A}(q, p_1, p_2)$. 

At tree level, this is the only contribution and already scales classically; therefore we can shift to special kinematics without worrying about the scaling,
\begin{equation}\label{treespinkick}
    \Delta^{(1)}S^{\mu \nu}_1 = -i \left[\delta^{(1)}_{\text{cov}}, S^{\mu \nu}_1\right]. 
\end{equation}
As a preliminary check, we can see whether Eq. (\ref{treespinkick}) satisfies spin-tensor-magnitude conservation
\begin{equation}\label{spinmagcons}
    2 S_{1 \, \mu \nu} \Delta S^{\mu \nu}_1 + \Delta S_{1\,\mu \nu}\Delta S_{1}^{\mu \nu} = 0,
\end{equation}
which is conserved regardless of whether we impose a SSC or not~\cite{Bern:2023ity, Alaverdian:2024spu}. Plugging in the eikonal phase related to our amplitude ansatz, Eq. (\ref{treelvlansatz}), we see that spin-tensor-magnitude conservation is satisfied 
\begin{equation}
    - 2 i S_{1 \, \mu \nu} \left[\delta^{(1)}_{\text{cov}}, S^{\mu \nu}_1\right] = 0,
\end{equation}
however this does not obviously vanish at the level of the eikonal phase. This signals that we are free to make a modification to the commutator that explicitly shows tree-level spin-tensor-magnitude conservation. In fact, one can find a projector for the commutator of the form
\begin{equation}\label{spinproj}
    \Sigma^{\mu \nu}_{1 \,\,\rho \sigma} = \delta^{\mu}_{\, \, \rho} \delta^{\nu}_{\, \sigma} - \frac{S^{\mu \nu}_1\, S_{1\,\rho \sigma}}{S_1^2},
\end{equation}
which leaves the commutator unchanged
\begin{equation}
    \left[S^{\mu \nu}_1, S^{\alpha \beta}_1\right] = \Sigma^{\mu \nu}_{1 \,\,\rho \sigma} \left[S^{\rho \sigma}_1, S^{\alpha \beta}_1\right] = \Sigma^{\alpha \beta}_{1 \,\,\gamma \delta} \left[S^{\mu \nu}_1, S^{\gamma \delta}_1\right] = \Sigma^{\mu \nu}_{1 \,\,\rho \sigma} \Sigma^{\alpha \beta}_{1 \,\,\gamma \delta} \left[S^{\rho \sigma}_1, S^{\gamma \delta}_1\right],
\end{equation}
while satisfying $\Sigma^{\mu \nu}_{1\,\,\rho \sigma} S^{\rho \sigma}_1 = 0$. In the same way as the momentum projector $\Pi^{\mu \nu}$ can be seen as a shift freedom associated with the on-shell $\delta$-functions~\cite{Luna:2023uwd}, the spin projector $\Sigma^{\mu \nu}_{1\,\,\rho \sigma}$ can be seen as a shift freedom associated with the Lorentz algebra. Both of these projectors are required to conserve their respective conservation conditions. We will see in the one-loop calculation that the spin projector will be just as crucial to calculating the correct spin kick as the momentum projector is to calculating the correct momentum impulse. 

Moving on to the one-loop correction, we must now decompose the amplitude into real and imaginary contributions in order to faithfully keep track of classical scaling. We will use a similar notation to that of the impulse case where
\begin{equation}
    \langle{\Psi}|  i [\mathbb{S}^{\mu \nu}_1, T] |{\Psi} \rangle|_{NLO}  = \Delta^{(2)}S^{\mu \nu}_{1 \, v, \text{Re}} + \Delta^{(2)}S^{\mu \nu}_{1 \, v, \text{Im}},    
\end{equation}
where 
\begin{equation}
    \Delta^{(2)}S^{\mu \nu}_{1 \, v, \text{Re}} =-i \int \cancel{D}q \, e^{- i b_{\text{cov}} \cdot q} \, \left[\text{Re}\mathcal{A}^{(2)}(q, p_1, p_2), S^{\mu \nu}_1 \right] = -i \left[\delta^{(2)}_{\text{cov}}, S^{\mu \nu}_1\right],
\end{equation}
and
\begin{equation}
    \Delta^{(2)}S^{\mu \nu}_{1 \, v, \text{Im}} = \int \cancel{D}q \, e^{- i b_{\text{cov}} \cdot q} \, \left[\text{Im}\mathcal{A}^{(2)}(q, p_1, p_2), S^{\mu \nu}_1 \right],
\end{equation}
where we will defer the expansion until later, when we consider the full iteration contribution. 

Plugging in the spin operator for the real kernel and fixing ourselves to one-loop order we get
\begin{align}
    \Delta^{(2)}S^{\mu \nu}_{1\, r} &\equiv \langle{\Psi}| \mathit{T}^{\dagger} \left[\mathbb{S}^{\mu \nu}_1, \mathit{T}\right] | \Psi \rangle  \nonumber \\
     &=  -\int \cancel{D}q e^{- i b \cdot q}\int \cancel{D}l \, \epsilon^{*} \cdot \mathbb{A}^{(1)}(q-l, p_1+l, p_2-l) \cdot 
    \left[\mathbb{A}^{(1)}(l, p_1, p_2), \mathit{M}^{\mu \nu} \right]\cdot \epsilon, 
\end{align}
which we will need to decompose into symmetric and antisymmetric parts, shift to special kinematics, and enforce the horizontal-flip symmetry.

\subsection{Iterative Contribution}
\hspace{0.6cm}We organize the iterative contribution in a similar manner as in the momentum impulse case,
\begin{equation}
    \Delta^{(2)}S^{\mu \nu}_{1\,it} \equiv \Delta^{(2)}S^{\mu \nu}_{1\, r} + \Delta^{(2)}S^{\mu \nu}_{1 \, v, \text{Im}} = \Delta^{(2)}S^{\mu \nu}_{1\, it, *} + \Delta^{(2)}S^{\mu \nu}_{1\, it, \partial u} + \Delta^{(2)}S^{\mu \nu}_{1\, it, \partial l}.
\end{equation}
Beginning with $\Delta^{(2)}S^{\mu \nu}_{1\, it, \partial u}$, this contribution is proportional to the external momentum derivatives 
\begin{align}\label{dSdu}
    \Delta^{(2)}S^{\mu \nu}_{1\, it, \partial u} \equiv&  -\frac{1}{2} \int \cancel{D}q e^{- i b \cdot q} \int \cancel{D}l \nonumber \\
    & \hspace{0.5cm}\epsilon^{*} \left(\left\{\left[\mathbb{A}^{(1)}(l, \bar{p}_1, \bar{p}_2), \mathit{M}^{\mu \nu} \right] , \frac{l}{2} \cdot \left(\frac{\partial}{\partial \bar{p}_1} - \frac{\partial}{\partial \bar{p}_2}\right) \mathbb{A}^{(1)}(q-l, \bar{p}_1, \bar{p}_2)\right\} \right. \nonumber \\ 
    & \left. \hspace{1.0cm}-\left\{\mathbb{A}^{(1)}(q-l, \bar{p}_1, \bar{p}_2), \left[ \frac{q-l}{2} \cdot \left(\frac{\partial}{\partial \bar{p}_1} - \frac{\partial}{\partial \bar{p}_2}\right) \mathbb{A}^{(1)}(l, \bar{p}_1, \bar{p}_2), \mathit{M}^{\mu \nu} \right]  \right\} \right) \epsilon.
\end{align}
Because Eq. (\ref{dSdu}) already scales classically and applying the horizontal-flip symmetry will not change the expression, we can immediately shift the momentum transfer and then express this contribution in terms of eikonal phases, 
\begin{align}
    \Delta^{(2)}S^{\mu \nu}_{1\, it, \partial u} =& \frac{i}{2} \left[\left(\frac{1}{\bar{m}_1}\frac{\partial}{ \partial u_1^{\alpha}} - \frac{1}{\bar{m}_2}\frac{\partial}{ \partial u_2^{\alpha}}\right) \delta^{(1)}_{\text{cov}}, S^{\mu \nu}_1 \right] \frac{\partial \delta^{(1)}_{\text{cov}}}{\partial (b_{\text{cov} \, \alpha})_{\perp}} \nonumber \\
    & -\frac{i}{2} \left[\frac{\partial \delta^{(1)}_{\text{cov}}}{\partial (b_{\text{cov} \, \alpha})_{\perp}}, S^{\mu \nu}_1 \right]\left(\frac{1}{\bar{m}_1}\frac{\partial}{ \partial u_1^{\alpha}}- \frac{1}{\bar{m}_2}\frac{\partial}{ \partial u_2^{\alpha}}\right) \delta^{(1)}_{\text{cov}} .
\end{align}

For the contribution proportional to the products of the amplitudes and the Lorentz generator
\begin{align}\label{dSprod}
    \Delta^{(2)}S^{\mu \nu}_{1\, it, *}  \equiv  &-\frac{1}{2} \int \cancel{D}q e^{- i b \cdot q} \int \cancel{D}l \nonumber \\
    &\hspace{0.75cm}\epsilon^{*}  \left(\left[ \mathbb{A}^{(1)}(q -l, \bar{p}_1, \bar{p}_2), \left[\mathbb{A}^{(1)}(l, \bar{p}_1, \bar{p}_2), \mathit{M}^{\mu \nu}\right]\right] \right. \nonumber\\
    & \left. \hspace{1.50cm}+ \left\{\mathbb{A}^{(1)}(q - l, \bar{p}_1, \bar{p}_2) ,\left[\mathbb{A}^{(1)}(l, \bar{p}_1, \bar{p}_2), \mathit{M}^{\mu \nu}\right]\right\} \right. \nonumber \\
    & \left. \hspace{2.0cm}- \frac{1}{2} \left[\left\{\mathbb{A}^{(1)}(q - l, \bar{p}_1, \bar{p_2}), \mathbb{A}^{(1)}(l, \bar{p}_1, \bar{p}_2)\right\}, \mathit{M}^{\mu \nu}\right]\right)\epsilon,
\end{align}
we still need to cancel the classically-singular contribution in the last two lines. When we examine these terms in Eq. (\ref{dSprod}) and apply commutator identities, we find that they are parity-odd under the horizontal-flip symmetry 
\begin{align}\label{dssuperclassicalcombo}
    &\left\{\mathbb{A}^{(1)}(q - l) ,\left[\mathbb{A}^{(1)}(l), \mathit{M}^{\mu \nu}\right]\right\}  - \frac{1}{2} \left[\left\{\mathbb{A}^{(1)}(q - l), \mathbb{A}^{(1)}(l)\right\}, \mathit{M}^{\mu \nu}\right] \nonumber \\
    & \hspace{1.0cm} = \frac{1}{2} \left(\left\{\mathbb{A}^{(1)}(q - l) ,\left[\mathbb{A}^{(1)}(l), \mathit{M}^{\mu \nu}\right]\right\} - \left\{\mathbb{A}^{(1)}(l) ,\left[\mathbb{A}^{(1)}(q-l), \mathit{M}^{\mu \nu}\right]\right\}\right),
\end{align}
where we leave the external momenta dependence implicit. Because the operator does not affect the parity of this contribution, the classically-singular terms vanish. Now that we have a fully classical contribution, we can follow the same procedure as before and rewrite Eq. (\ref{dSprod}) in terms of tree-level eikonal phases
\begin{equation}
    \Delta^{(2)}S^{\mu \nu}_{1\, it, *} = -\frac{1}{2} \left[\delta^{(1)}_{\text{cov}}, \left[\delta^{(1)}_{\text{cov}}, S^{\mu \nu}_1\right]\right]. 
\end{equation}

For $\Delta^{(2)}S^{\mu \nu}_{1\, it, \partial l}$, we first need to integrate by parts in order to apply the horizontal-flip symmetry. Doing so results in the same cancellation of classically-singular terms as in $\Delta^{(2)}S^{\mu \nu}_{1\, it, *}$, where using Eq. (\ref{dssuperclassicalcombo}) results in a parity-odd contribution. However, we run in to the same obstacle as in the momentum impulse case,
\begin{align}
    \Delta^{(2)}S^{\mu \nu}_{1\, it, \partial l} =   &-\int \cancel{D}q e^{- i b \cdot q} \int \cancel{D}l \frac{-q^2}{4}  \left(\frac{\check{u}^{\alpha}_2}{\bar{m}_2} - \frac{\check{u}^{\alpha}_1}{\bar{m}_1}\right)\nonumber \\
    & \hspace{0.75cm} \left(\mathcal{A}^{(1)}(q-l) \left[ \frac{\partial}{\partial l^{\alpha}}\mathcal{A}^{(1)}(l), S^{\mu \nu}_1 \right] - \frac{\partial}{\partial l^{\alpha}}\mathcal{A}^{(1)}(l) \left[\mathcal{A}^{(1)}(q-l), S^{\mu \nu}_1\right] \right), 
\end{align}
which is overall parity-even and scales classically. As before, we have to resolve the $\partial/\partial l^{\alpha}$ derivative; luckily the combination is the same as in the momentum impulse case. Therefore, we can perform the transfer momentum shift and use the same replacement rule Eq. (\ref{replacementruledl}) to express $\Delta^{(2)}S^{\mu \nu}_{1\, it, \partial l}$ in terms of eikonal phases
\begin{equation}
    \Delta^{(2)}S^{\mu \nu}_{1\, it, \partial l} = - i \left(\frac{\check{u}_1^{\alpha}}{\bar{m}_1} - \frac{\check{u}_2^{\alpha}}{\bar{m}_2}\right) \left(  \left[ \frac{\partial \delta^{(1)}_{\text{cov}}}{\partial \Pi^{\alpha \gamma}}, S^{\mu \nu}_1 \right] \frac{\partial \delta^{(1)}_{\text{cov}}}{\partial (b_{\text{cov}\,\gamma})_{\perp}}  -    \left[ \frac{\partial \delta^{(1)}_{\text{cov}}}{\partial (b_{\text{cov}\,\gamma})_{\perp}}, S^{\mu \nu}_1 \right] \frac{\partial\delta^{(1)}_{\text{cov}}}{\partial \Pi^{\alpha \gamma}} \right).  
\end{equation}
With the full iterative contribution in terms of eikonal phases, we now have the higher-order-in-spin eikonal formula to one-loop order for the spin kick. However, we would like to first make one-loop spin-tensor-magnitude conservation explicit by incorporating the projector Eq. (\ref{spinproj}) associated with the Lorentz algebra into our commutators. For most of the contributions no appreciable difference occurs, but in the case of $\Delta^{(2)}S^{\mu \nu}_{1\, it, *}$ the double commutator has a non-trivial effect
\begin{equation}
    \Delta^{(2)}S^{\mu \nu}_{1\, it, *} = -\frac{1}{2} \Sigma^{\mu \nu}_{1\,\,\rho \sigma} \left[\delta^{(1)}_{\text{cov}}, \left[\delta^{(1)}_{\text{cov}}, S^{\rho \sigma}_1\right]\right] + S^{\mu \nu}_1 \frac{\left[\delta^{(1)}_{\text{cov}}, S^{\rho \sigma}_1\right]\left[\delta^{(1)}_{\text{cov}}, S_{1\,\rho \sigma}\right]}{2 S^2_1}.
\end{equation}
Now we can combine our iteration terms knowing that our eikonal formula for the spin kick will explicitly satisfy one-loop spin-tensor-magnitude conservation
\begin{align}\label{spinkick}
     \Delta^{(2)}S^{\mu \nu}_{1} = &\, -i \left[\delta^{(2)}_{\text{cov}}, S^{\mu \nu}_1\right] - \frac{1}{2} \Sigma^{\mu \nu}_{1\,\,\rho \sigma} \left[\delta^{(1)}_{\text{cov}}, \left[\delta^{(1)}_{\text{cov}}, S_1^{\rho \sigma} \right]\right] + \frac{1}{2} S^{\mu \nu}_1 \frac{\left[\delta^{(1)}_{\text{cov}}, S^{\rho \sigma}_1\right]\left[\delta^{(1)}_{\text{cov}}, S_{1\,\rho \sigma}\right]}{S^2_1}\nonumber \\
     & - \frac{i}{2} \left[\frac{\partial \delta^{(1)}_{\text{cov}}}{\partial (b^{\alpha}_{\text{cov}})_{\perp}}, S^{\mu \nu}_1\right]  \nabla^{\alpha}_{p_{\text{cm}}} \delta^{(1)}_{\text{cov}}  + \frac{i}{2}  \frac{\partial \delta^{(1)}_{\text{cov}}}{\partial (b^{\alpha}_{\text{cov}})_{\perp}}  \nabla^{\alpha}_{p_{\text{cm}}} \left[\delta^{(1)}_{\text{cov}}, S^{\mu \nu}_1\right].
\end{align}
We have verified Eq. (\ref{spinkick}) to the results in Ref.~\cite{Luna:2023uwd} up to quadratic order in spin. We have also verified that Eq. (\ref{spinkick}) does respect spin-tensor-magnitude conservation. 

Once again we can rewrite Eq. (\ref{spinkick}) in terms of tree-level spin kicks and momentum impulses
\begin{align}\label{spinkickItForm}
     \Delta^{(2)}S^{\mu \nu}_{1} = &\, -i \left[\delta^{(2)}_{\text{cov}}, S^{\mu \nu}_1\right] - \frac{i}{2} \Sigma^{\mu \nu}_{1\,\,\rho \sigma} \left[\delta^{(1)}_{\text{cov}}, \Delta^{(1)}S^{\rho \sigma}_1\right] -  \frac{S^{\mu \nu}_1}{2 S^2_1} (\Delta^{(1)}S_1)^2 \nonumber \\
     & \,\,\,+ \frac{i}{2} \left[\Delta^{(1)}p_{1\,\alpha}, S^{\mu \nu}_1\right]  \nabla^{\alpha}_{p_{\text{cm}}} \delta^{(1)}_{\text{cov}}  - \frac{i}{2} \Delta^{(1)}p_{1\,\alpha}  \left[\nabla^{\alpha}_{p_{\text{cm}}} \delta^{(1)}_{\text{cov}}, S^{\mu \nu}_1\right],
\end{align}
which bears a striking resemblance to Eq. (\ref{dPimpform}). We also find that the cut-correction effects vanish when we impose a SSC. 

\section{Comparing the Eikonal Formulas}
\hspace{0.6cm}As can be seen in Eq. (\ref{dPimpform}) and Eq. (\ref{spinkickItForm}), after taking into account the conservation equations for the momentum and the spin tensor, both the impulse and spin kick formulas exhibit the following pattern
\begin{align}\label{patternansatz}
    \Delta^{(2)}\mathcal{O} = & \mathbb{O} \circ \delta^{(2)}_{\text{cov}} - \frac{i}{2}  \left[\delta^{(1)}_{\text{cov}},  \mathbb{O}\circ \delta^{(1)}_{\text{cov}} \right]  - \frac{1}{2} \mathbb{O} \overleftrightarrow{\circ} \left(\Delta^{(1)} p_{1\,\alpha} \, \nabla^{\alpha}_{p_{\text{cm}}} \delta^{(1)}_{\text{cov}} \right),
\end{align}
where $\Delta \mathcal{O}$ can be either observable $\Delta p_1^{\mu}, \Delta S^{\mu \nu}_1$ and $\mathbb{O}$ is the operator associated with the observable such that 
\begin{equation}
    \mathbb{S}^{\mu \nu}_1 \circ \delta \equiv - i \left[\delta, S^{\mu \nu}_1\right], \hspace{3cm} \mathbb{P}^{\mu}_1 \circ \delta \equiv -\frac{\partial \delta}{\partial (b_{\text{cov} \, \mu})_{\perp}}.
\end{equation}
 We define the operation $\overleftrightarrow{\circ}$ as
\begin{equation}
    \mathbb{O} \overleftrightarrow{\circ} \left(\nabla^{\alpha}_{p_{\text{cm}}} \delta^{(1)}_{\text{cov}} \, \Delta^{(1)} p_{1\,\alpha}\right) \equiv \left(\mathbb{O} \circ\nabla^{\alpha}_{p_{\text{cm}}}  \delta^{(1)}_{\text{cov}} \right) \, \Delta^{(1)} p_{1\,\alpha} -\nabla^{\alpha}_{p_{\text{cm}}}  \delta^{(1)}_{\text{cov}} \, \left( \mathbb{O} \circ\Delta^{(1)} p_{1\,\alpha}\right).
\end{equation}
While it is unclear where such a pattern would arise from, there have been suggestions in the literature of sources for eikonal formulas. In Ref.~\cite{Bern:2020buy}, the authors suggest that their eikonal formula arises from an exponential operation on the observables of the following form 
\begin{equation}\label{HamEik}
    \Delta \mathcal{O} = e^{- i \delta \mathcal{D}} \left[\mathcal{O}, e^{ i \delta \mathcal{D}}\right],
\end{equation}
where $\mathcal{D}$ is some differential operator, which is defined in Eq. (7.21) of Ref.~\cite{Bern:2020buy}. In Refs.~\cite{Luna:2023uwd, Jakobsen:2021zvh}, both works show that the higher-order corrections to observables can be generated by making half variable shifts to their observables
\begin{align}
    \Delta \mathcal{O} (u_i, s_i, b) \rightarrow \Delta \mathcal{O} \left(u_i + \frac{\Delta u_i}{2}, s_i + \frac{\Delta s_i}{2}, b + \frac{\Delta b}{2} \right),
\end{align}
 where expanding to next to leading order produces their one-loop corrections, which we agree with. It is clear that if we interpret the exponential operator in Eq. (\ref{HamEik}) as some translation operators that these two formulas are related. In fact, Ref.~\cite{Luna:2023uwd} showed that the eikonal formulas in Refs.~\cite{Bern:2020buy, Bern:2023ity}, with the SSC freedom, can be expressed as directional derivatives, which would come from expanding a small shift on the variables of the observables. We believe that our formulas can also be expressed in such a form, however, an explicit calculation still needs to be done.

\section{Conclusion}

\hspace{0.6cm}In this paper we used the KMOC formalism to derive one-loop formulas that relate the eikonal phase to the momentum impulse and the spin kick with spin effects included. We have verified these equations up to quadratic order in spin by comparing to Ref.~\cite{Luna:2023uwd}. We found that our eikonal formulas, Eq. (\ref{dPimpform}) and Eq. (\ref{spinkickItForm}), follow a Baker-Campbell-Hausdorff expansion pattern, Eq. (\ref{patternansatz}), similar to those found in Refs.~\cite{Bern:2020buy, Alaverdian:2024spu, Bern:2023ity, Kosmopoulos:2021zoq}; Ref.~\cite{Luna:2023uwd} found this pattern is related to half-variable shifts in the observables, signaling that the eikonal formulas can be generated by some translation operator on the observables. We also found that it was crucial to include projectors Eq. (\ref{projector}) and Eq. (\ref{spinproj}), which are associated with the conservation of energy and spin tensor magnitude, respectively. In particular, we found that the projector Eq. (\ref{spinproj}) leaves the Lorentz algebra unchanged, and therefore is a projective freedom associated with it. 

We chose to use non-transverse massive spinning fields in order to capture SSC-violating effects, such as spin vector magnitude violation, with the understanding that we can later impose the SSC to match results using transverse fixed-spin fields~\cite{Bern:2023ity, Alaverdian:2024spu}: this simplified our calculations considerably. As a result, we were able to use simplified completeness relations for the polarization tensors of the massive spinning fields; this allowed for the easy manipulation of integrands proportional to the on-shell product of tree-level amplitudes. We were also able to assume an analytic structure of our tree-level amplitudes; we did this in order to resolve the cut-correction effect in Eq. (\ref{troubletermdp}) that resulted from expanding the on-shell $\delta$-functions of the two-particle cut integral associated with the iteration contributions. We found that, upon imposing a SSC, this effect no longer contributes to our observables. This relationship works well with our understanding that, by using SSC-violating fields, we allow for lower spin states to propagate in our amplitudes, which would be eliminated by enforcing a SSC. Since these cut-correction contributions come from the unitarity cut of the internal massive lines, and imposing the SSC removes them, we believe they may be shown to be related to the effects of the lower spin states propagating in our amplitudes, such as the spin vector magnitude change. It would be interesting to see an explicit calculation showing the relationship between the cut corrections and the spin vector magnitude change.   

A number of avenues exist to expand on this work. Because we checked our equations up to quadratic order in spin, we were not able to see the subtleties associated with higher-spin amplitudes starting at quintic order in spin~\cite{Bautista:2022wjf}. Though, in principle, our derivation is valid to any order in spin, it would be important to check how these higher-spin subtleties show up in the context of this work. Another extension would be to see how to derive an eikonal formula at two loops, especially since the iteration pieces in this order will also include one-loop contributions as well as on-shell massless particles for the tree-level iterations. It would also be interesting to see how non-conservative effects, such as radiation and absorption, are affected by spin degrees of freedom. Recently, work in Refs.~\cite{Aoude:2024jxd, Bautista:2024emt} have explored the effects of spin on mass absorption and the momentum impulse. Exploring how absorption affects the spin kick would also be insightful, particularly in understanding how spin transitions play a role. Furthermore, it would be interesting to see shifts like those in Ref.~\cite{Jakobsen:2021zvh}, which preserve the symmetry in the worldline~\cite{Luna:2023uwd}, affect the eikonal formulas in a similar way as the momentum-conserving projector Eq. (\ref{projector}) or the spin-tensor-magnitude-conserving projector Eq. (\ref{spinproj}).

\section*{Acknowledgements}
We would like to thank Zvi Bern, Andres Luna, Trevor Scheopner, Callum Jones, Michael Ruf, Richard Myers, Justin Vines, Fei Teng, Giulia Isabella, and Rafael Aoude for useful discussions that were paramount to this work. This work was supported in part by the U.S. Department of Energy (DOE) under Award Number
DE-SC0009937. We are also grateful for support from the Mani L. Bhaumik Institute for Theoretical Physics.

\newpage

\appendix

\section{Elements for Comparison}
To compare our eikonal formulas to Ref.~\cite{Luna:2023uwd}, we matched our amplitudes to theirs while imposing the SSC and changing to their basis of variables. We then calculated the eikonal phases for these amplitudes and plugged these into the eikonal formulas and compared observables. 

Because Ref.~\cite{Luna:2023uwd} used the spin vector as their spin variable, we needed to use the relationship,
\begin{equation}\label{tensortovecnoK}
    S^{\mu \nu}(p) = \frac{1}{m} \epsilon^{\mu \nu \rho \sigma} p_{\rho} s_{\sigma}(p).
\end{equation}
Note that we could have included the mass dipole $K^{\mu}$ in Refs.~\cite{Alaverdian:2024spu, Bern:2023ity}, but because we will be imposing the SSC this would be pointless.

We used the following coefficients for our amplitudes Eq. (\ref{treelvlansatz}) and Eq. (\ref{2pmansatz}):
\setlength{\jot}{7pt}
\begin{align}
    &\Upsilon^{(1)} = c^{(1, 0, 0)} ,\\
    & \Upsilon^{(1)}_{\nu_1}  = c^{(1, 1, 0)}_{1} u_{1\,\nu_1} + c^{(1, 1, 0)}_{2} u_{2 \, \nu_1} ,\\
    &\Upsilon^{(1)}_{\nu_1 \nu_2}  =  c^{(1, 2, 0)}_{1} \eta_{\nu_1 \nu_2} + c^{(1, 2, 0)}_{2} u_{1 \, \nu_1} u_{1 \, \nu_2} + c^{(1, 2, 0)}_{3} u_{2 \, \nu_1} u_{2 \, \nu_2} + c^{(1, 2, 0)}_{4} \left(u_{1 \, \nu_1} u_{2 \, \nu_2} + u_{2 \, \nu_1} u_{1 \, \nu_2}\right) ,\\
    &\Upsilon^{(2)} = c^{(2, 0, 0)} ,\\
    &\Upsilon^{(2)}_{\nu_1}  = c^{(2, 1, 0)}_{1} u_{1\,\nu_1} + c^{(2, 1, 0)}_{2} u_{2 \, \nu_1} ,\\
    &\Upsilon^{(2)}_{\nu_1 \nu_2}  =  c^{(2, 2, 0)}_{1} \eta_{\nu_1 \nu_2} + c^{(2, 2, 0)}_{2} u_{1 \, \nu_1} u_{1 \, \nu_2} + c^{(2, 2, 0)}_{3} u_{2 \, \nu_1} u_{2 \, \nu_2} + c^{(2, 2, 0)}_{4} \left(u_{1 \, \nu_1} u_{2 \, \nu_2} + u_{2 \, \nu_1} u_{1 \, \nu_2}\right),\\
    &\Upsilon^{(2)}_{\mu_1 \nu_1 \mu_2 \nu_2} = c^{(2, 2, 0)}_{q^2, 1} \left(\eta_{\mu_1 \mu_2}\eta_{\nu_1 \nu_2} - \eta_{\nu_1 \mu_2}\eta_{\mu_1 \nu_2}\right) \nonumber \\
    &\hspace{1.9cm}+ c^{(2, 2, 0)}_{q^2, 2}\left(\eta_{\mu_1 \mu_2} u_{1\,\nu_1}u_{1\,\nu_2} + \eta_{\nu_1 \nu_2} u_{1\,\mu_1}u_{1\,\mu_2}  - \eta_{\nu_1 \mu_2} u_{1\,\mu_1}u_{1\,\nu_2} - \eta_{\mu_1 \nu_2} u_{1\,\nu_1}u_{1\,\mu_2}\right) \nonumber \\
    &\hspace{1.9cm}+ c^{(2, 2, 0)}_{q^2, 3}\left(\eta_{\mu_1 \mu_2} u_{2\,\nu_1}u_{2\,\nu_2} + \eta_{\nu_1 \nu_2} u_{2\,\mu_1}u_{2\,\mu_2} - \eta_{\nu_1 \mu_2} u_{2\,\mu_1}u_{2\,\nu_2} - \eta_{\mu_1 \nu_2} u_{2\,\nu_1}u_{2\,\mu_2}\right) \nonumber \\
    & \hspace{1.9cm}+ c^{(2, 2, 0)}_{q^2, 4} \left(\eta_{\mu_1 \mu_2} \left(u_{1\,\nu_1}u_{2\,\nu_2} + u_{2\,\nu_1}u_{1\,\nu_2}\right) + \eta_{\nu_1 \nu_2} \left(u_{1\,\mu_1}u_{2\,\mu_2} + u_{2\,\mu_1}u_{1\,\mu_2}\right) \right. \nonumber \\
    &\left. \hspace{3.6cm} - \eta_{\nu_1 \mu_2} \left(u_{1 \, \mu_1} u_{2 \, \nu_2} + u_{2 \, \mu_1} u_{1 \, \nu_2}\right) - \eta_{\mu_1 \nu_2} \left(u_{1 \, \nu_1} u_{2 \, \mu_2} + u_{2 \, \nu_1} u_{1 \, \mu_2}\right)\right) \nonumber\\
    & \hspace{1.9cm}+ c^{(2, 2, 0)}_{q^2, 5} \left(u_{1 \, \mu_1} u_{1\,\mu_2}u_{2 \, \nu_1} u_{2\,\nu_2} + u_{2 \, \mu_1} u_{2 \, \mu_2} u_{1 \, \nu_1} u_{1 \, \nu_2} \right. \nonumber \\
    & \left. \hspace{3.6cm} - u_{1 \, \nu_1} u_{1\,\mu_2}u_{2 \, \mu_1} u_{2\,\nu_2}- u_{1 \, \mu_1} u_{1\,\nu_2}u_{2 \, \nu_1} u_{2\,\mu_2}\right),
\end{align}
where the superscript in the coefficients follow the pattern $c^{(\text{nPM}, \, s_1, \, s_2)}$ and the $c^{(2, 2, 0)}_{q^2, i}$ come from the extra contributions that appear starting at 2PM. The tensor structures above were chosen such that they followed the symmetries of the spin tensors and their products. 

To properly compare amplitudes, we need to enforce the covariant SSC,
\begin{equation}
    p_{1 \, \mu}S^{\mu \nu}_1 = 0,
\end{equation}
on our coefficients. To do this, we perturbatively expand the SSC, 
\begin{equation}
    \left( p_{1\,\mu} + \Delta p_{1\,\mu} \right)\left( S^{\mu \nu}_1 + \Delta S^{\mu \nu}_1 \right) = 0,
\end{equation}
and treat each order in PM and spin as an equation of constraint on our coefficients. After enforcing these constraints, and switching to the basis $\left\{q, u_1, u_2, \epsilon^{\mu \nu \rho \sigma} u_{1 \, \nu} u_{2 \, \rho} q_{\sigma}\right\}$ we can match our coefficients to those of the amplitudes in Ref.~\cite{Luna:2023uwd}. In principle, we can use this procedure to match to any amplitude. 

Now that we are effectively using the same amplitudes, we can Fourier transform our amplitudes into eikonal phases using the following integral
\begin{equation}
    \int \hat{d}^{D}q \, \hat{\delta}\left(2 \bar{p}_1 \cdot q \right) \hat{\delta}\left(2 \bar{p}_2 \cdot q \right) \frac{e^{- i q \cdot b_{\text{cov}}}}{\left(-q^2\right)^{\alpha}} = \frac{1}{4 \bar{m}_1\bar{m}_2 \sqrt{y^2 - 1}}\frac{\Gamma(D/2 - 1 - \alpha)}{2^{2\alpha} \pi^{(D/2 - 1)} \Gamma(\alpha)} \frac{1}{\left(-b_{\text{cov}}^2\right)^{(D/2 - 1 - \alpha)}},
\end{equation}
where we exchange any factors of $i q^{\mu}$ in the amplitude for $\Pi^{\mu \nu} \partial/\partial b_{\text{cov}}^{\nu}$. 

To compare the spin kick effect, we need to invert Eq. (\ref{tensortovecnoK}) to solve for $s^{\mu}$ and then perturbatively expand,
\begin{equation}
    \Delta s^{\mu}_1 = \frac{1}{2 m_1} \epsilon^{\mu \nu \rho \sigma} \left(\Delta p_{1 \, \nu} S_{1\,\rho \sigma} + p_{1 \, \nu}  \Delta  S_{1\,\rho \sigma} + \Delta p_{1 \, \nu}  \Delta  S_{1\,\rho \sigma}\right). 
\end{equation}
At this point, we have all we need to compare our formulas to the results of Ref.~\cite{Luna:2023uwd}, which we find agreement with.

\section{Polarization Tensors in Special Kinematics}
In principle, we should consider how shifting to special kinematics would affect our polarizations, and therefore the spin tensor. However, as we see in the derivation, this effect does not affect our calculation due to reasons of horizontal-flip symmetry and classical counting.

When shifting Eq. (\ref{polexp}) to special kinematics, we arrive at the simple replacement $\omega_i(p_i)\rightarrow - \omega_i(\bar{p}_i) \equiv - \bar{\omega}_i$. Therefore the effect on $b_{\text{cov}}$ can be taken into account entirely by the replacement $b_{\text{cov}} \rightarrow \bar{b}_{\text{cov}} = b + \left(\bar{\omega}_1 - \bar{\omega}_2\right)$.

The effect on the spin tensor is more complicated. Starting with $S^{\mu \nu}(p_1) \rightarrow S^{\mu \nu}(\bar{p}_1 -q/2)$, which occurs when we shift to special kinematics, but in terms of polarization tensors, we see that
\begin{align}
    \epsilon^{*}(\bar{p}_1 - q/2) & \cdot \mathit{M} \cdot  \epsilon(\bar{p}_1 - q/2) \nonumber \\
    &= \left(\epsilon^{*}(\bar{p}_1 - q/2) \cdot \epsilon(\bar{p}_1)\right)\left( \epsilon^{*}(\bar{p}_1) \cdot \mathit{M} \cdot  \epsilon(\bar{p}_1) \right)\left( \epsilon^{*}(\bar{p}_1) \cdot \epsilon(\bar{p}_1 - q/2)\right) \nonumber \\
    & = \epsilon^{*}(\bar{p}_1) \cdot e^{i q \cdot \bar{\omega}_1/2} \cdot \mathit{M} \cdot  e^{-i q \cdot \bar{\omega}_1/2} \cdot \epsilon(\bar{p}_1) \nonumber \\
    & = S^{\mu \nu }(\bar{p}_1) + \frac{1}{2} \left[i q \cdot \bar{\omega}_1 ,  S^{\mu \nu}(\bar{p}_1) \right] + \cdots \label{sqcorrection}
\end{align}
where in the last line we expanded using the identity due to Baker, Campbell, and Hausdorff and truncated the expansion to leading order in $q$. The second term scales higher in classical scaling and therefore would be something to take into account when we keep track of classical contributions. Note that this correction results in the same one for the spin vector in Refs.~\cite{Maybee:2019jus, Cristofoli:2021jas, Luna:2023uwd} if we choose the reference momentum to be $k_i = \bar{p}_i$. 

In our set up, we never have to include internal polarization tensors because these will just be turned to Kronecker deltas. Therefore the only correction relating to the shifted spin tensor we ever have to consider are proportional to $q$, which does not affect the horizontal-flip symmetry parity. Because the parity is unchanged when we use this effect to promote our classically-singular terms, we will find the odd parity of these terms will make them vanish. As a result, we can conclude that the special kinematics shift to the spin tensor does not affect our calculation, unlike in the fixed-spin formalism of Refs.~\cite{Maybee:2019jus, Cristofoli:2021jas, Luna:2023uwd}.

\section{Resolving the Cut-Correction Term}
Here we present how we arrive at the cut-correction contribution to our observables. We will first plug in the ansatz for our tree amplitudes in Eq. (\ref{treelvlansatz}) to Eq. (\ref{troubletermdp}) and then compare this to a directional derivative with respect to the projector $\Pi^{\mu}_{\,\, \nu}$. We will be simplifying certain aspects of the integrand in the observables analysis without affecting the basic conclusions. 

Plugging in our ansatz to the simplified integrand results in the following,
\begin{align}\label{divexplicit}
   & \int \cancel{D}l e^{- i b \cdot l}\frac{u^{\gamma}}{m} \Delta p_{\alpha} l^{\alpha}\frac{\partial \mathcal{A}^{(1)}(l)}{\partial l^{\gamma}} \nonumber \\
   &= \Delta p_{\alpha} \frac{u^{\gamma}}{m} (-1)^{(s_1 + s_2)} \left(s_1 \eta_{\gamma \mu_{s_1}} \Pi_{\rho_{s_2} \beta} + s_2 \eta_{\rho_{s_2}\gamma} \Pi_{\mu_{s_1}\beta}\right)\frac{\partial}{\partial b_{\alpha}}\frac{\partial}{\partial b_{\beta}}\frac{\partial}{\partial b^{\mu(s_1 - 1)}_{\perp}}\frac{\partial}{\partial b^{\rho(s_2 - 1)}_{\perp}}f^{\mu(s_1)\rho(s_2)}_{b^2},
\end{align}
 where we absorb irrelevant terms into the function $f^{\mu(s_1)\rho(s_2)}_{b^2}$. Note that we drop all terms proportional to $u\cdot b = 0$. Eq. (\ref{divexplicit}) provides a clue of where to go: it would appear that we have replaced the original projector dressing the impact parameter derivatives with a new projector via some product rule. Let us now consider the following directional derivative 
\begin{align}
    &\frac{u^{\gamma}}{m} \Delta p^{\alpha} \frac{\partial \delta^{(1)}}{\partial \Pi^{\gamma \alpha}} = \frac{u^{\gamma}}{m} \Delta p^{\alpha} (-1)^{(s_1 + s_2)} \frac{\partial }{\partial \Pi^{\gamma \alpha}} \left(\Pi_{\mu(s_1)\nu(s_1)} \Pi_{\rho(s_2)\sigma(s_2)}\right) \frac{\partial}{\partial b_{\nu(s_1)}}\frac{\partial}{\partial b_{\sigma(s_2)}} f^{\mu(s_1)\rho(s_2)}_{b^2} \nonumber \\
    &= (-1)^{(s_1 + s_2)}\frac{u^{\gamma}}{2 m} \Delta p_{\alpha}\left(s_1 \eta_{\gamma \mu_{s_1}} \Pi_{\rho_{s_2} \beta} + s_2 \eta_{\rho_{s_2} \gamma} \Pi_{\mu_{s_1} \beta}\right) \frac{\partial}{\partial b_{\alpha}}\frac{\partial}{\partial b_{\beta}}\frac{\partial}{\partial b^{\mu(s_1-1)}_{\perp}}\frac{\partial}{\partial b^{\rho(s_2-1)}_{\perp}}f^{\mu(s_1)\rho(s_2)}_{b^2} \nonumber \\
    & \hspace{2cm} = \frac{1}{2} \int \cancel{D}l e^{- i b \cdot l}\frac{u^{\gamma}}{m} \Delta p_{\alpha} l^{\alpha}\frac{\partial \mathcal{A}^{(1)}(l)}{\partial l^{\gamma}},
\end{align} 
where we define the derivative
\begin{equation}
    \frac{\partial \Pi_{\mu \nu}}{\partial \Pi_{\alpha \beta}} \equiv \frac{1}{2} \left(\delta^{\alpha}_{\mu} \delta^{\beta}_{\nu} + \delta^{\beta}_{\mu} \eta^{\alpha}_{\nu}\right),
\end{equation}
based on the symmetric nature of the projector. This is more of a bookkeeping notation rather than literally taking a derivative by a projector, which may prove to be a mathematically heretical thing to do. We could have obtained a similar result if we instead did not treat the projector as an independent variable and rather as a function of the velocities $u_i$, 
\begin{equation}\label{dduproj}
    \Delta p \cdot \frac{\partial \delta^{(1)}}{\partial u_i}|_{\frac{\partial \Upsilon}{\partial u_i} \rightarrow 0} =  - 2 \Delta p^{\alpha} \check{u}^{\gamma}_i \frac{\partial \delta^{(1)}}{\partial \Pi^{\alpha \gamma}},
\end{equation}
where we have isolated derivatives acting on the projectors. We will consider the derivative with respect to the projector as a sort of shorthand for the left-hand side of Eq. (\ref{dduproj}). 

\newpage

\bibliography{jheprefs.bib}

\end{document}